\documentclass[reprint,superscriptaddress,amsmath,amssymb,aps,showpacs]{revtex4-1}
\usepackage{hyperref}
\usepackage{comment}
\usepackage{graphicx}
\usepackage{color}

\hypersetup{
colorlinks = true,
linkcolor = blue,
citecolor = blue,
urlcolor = blue
}

\newcommand{\up}{\uparrow}
\newcommand{\dn}{\downarrow}
\newcommand{\jpot}{\frac{J_\perp}{t}}
\newcommand{\Up}{\Uparrow}
\newcommand{\Dn}{\Downarrow}
\newcommand{\ee}{\t{e}}
\newcommand{\ii}{\t{i}}
\renewcommand{\t}[1]{\text{#1}}
\newcommand{\eref}[1]{Eq.~(\ref{#1})}
\newcommand{\cross}{\times}
\newcommand{\dagg}[1]{#1^{\dag}}
\newcommand{\lp}{\left(}
\newcommand{\rp}{\right)}
\newcommand{\lb}{\left[}
\newcommand{\rb}{\right]}
\newcommand{\beqn}{\begin{eqnarray*} }
\newcommand{\eeqn}{\end{eqnarray*} }
\newcommand{\beqnn}{\begin{eqnarray} }
\newcommand{\eeqnn}{\end{eqnarray} }
\newcommand{\beq}{\begin{equation} }
\newcommand{\eeq}{\end{equation} }
\newcommand{\benum}{\begin{enumerate} }
\newcommand{\eenum}{\end{enumerate} }
\newcommand{\bbmat}{\begin{bmatrix} }
\newcommand{\ebmat}{\end{bmatrix} }
\newcommand{\bvmat}{\begin{vmatrix} }
\newcommand{\evmat}{\end{vmatrix} }
\newcommand{\e}{&=&}

\newcommand{\ds}{\displaystyle}

\newcommand{\non}{\nonumber}
\renewcommand{\v}[1]{\mathbf{#1}}
\newcommand{\rf}[1]{\mathbf{\hat{#1}}}

\newcommand{\fref}[1]{Fig.~\ref{#1}}
\newcommand{\tref}[1]{Tab.~\ref{#1}}
\newcommand{\bra}[1]{\left\langle#1\right|}
\newcommand{\ket}[1]{\left|#1\right\rangle}
\newcommand{\braket}[2]{\langle#1|#2\rangle}
\newcommand{\Braket}[3]{\left\langle#1\left|#2\right|#3\right\rangle}
\newcommand{\ev}[1]{\left\langle#1\right\rangle}
\newcommand{\bc}{\begin{center} }
\newcommand{\ec}{\end{center} }
\newcommand{\abs}[1]{\left|#1\right|}
\renewcommand{\star}[1]{#1^{\ast}}

\begin{document}

\title{$d$-Wave Superfluidity in Optical Lattices of Ultracold Polar Molecules}

\author{Kevin A. Kuns}
\affiliation{Institute for Quantum Information, California Institute of Technology, Pasadena, California 91125, USA}

\author{Ana Maria Rey}
\affiliation{JILA, National Institute of Standards and Technology and University of Colorado, Boulder, Colorado 80309-0440\\
and Department of Physics, University of Colorado, Boulder, Colorado 80309-0390, USA}

\author{Alexey V. Gorshkov}
\affiliation{Institute for Quantum Information, California Institute of Technology, Pasadena, California 91125, USA}

\date{\today}

\begin{abstract}
Recent work on ultracold polar molecules, governed by a generalization of the $t$-$J$ Hamiltonian, suggests that molecules may be better suited than atoms for studying $d$-wave superfluidity due to stronger interactions and larger tunability of the system. We compute the phase diagram for polar molecules in a checkerboard lattice consisting of weakly coupled square plaquettes. In the simplest experimentally realizable case where there is only tunneling and an $XX$-type spin-spin interaction, we identify the parameter regime where $d$-wave superfluidity occurs. We also find that the inclusion of a density-density interaction destroys the superfluid phase and that the inclusion of a spin-density or an Ising-type spin-spin interaction can enhance the superfluid phase. We also propose schemes for experimentally realizing the perturbative calculations exhibiting enhanced $d$-wave superfluidity.
\end{abstract}


\maketitle

\section{Introduction}

The Hubbard Hamiltonian is believed to contain some of the ingredients necessary to explain high-temperature superconductivity in cuprates \cite{Lee2006,Ogata2008,Dagotto1994}. The difficulties of analytically understanding the Hubbard Hamiltonian in more than one dimension suggest  the use of experimental quantum simulators to investigate the physics of this model, and important experimental progress in this direction has been made with lattices of ultracold atoms \cite{Schneider2008,Jordens2008,Simon2011,Sherson2010,Bloch2008,Trotzky2008,Trotzky2010}. In the context of high-temperature superconductivity, the most relevant regime of the Hubbard Hamiltonian is the limit of strong on-site interactions, in which the model reduces to the $t$-$J$ Hamiltonian \cite{Lee2006,Ogata2008,Dagotto1994}. Here $t$ is the nearest-neighbor tunneling amplitude and $J$ is the nearest-neighbor exchange interaction originating from second-order virtual hopping. Unfortunately, the exchange interactions are so small that it is extremely difficult to observe the associated physics in the cold atom implementation \cite{Bloch2008}.

Recently, the polar molecules KRb and LiCs have been cooled to their electronic, rotational, and vibrational ground states \cite{Carr2009,Ni2008,Aikawa2010,Deiglmayr2008}, and KRb has been loaded into a three-dimensional optical lattice \cite{Chotia2011}. This system can be used to implement lattice Hamiltonians based on rotational states of polar molecules \cite{Barnett2006,Micheli2006,Brennen2007,Buchler2007,Watanabe2009,Wall2009,Yu2009,Krems2009,Wall2010,Schachenmayer2010,Perez-Rios2010,Trefzger2010,Herrera2010,Kestner2011,Gorshkov2011,Gorshkov2011b,Liu2011}. Specifically, two rotational states of the molecules can be used as an effective spin-1/2 degree of freedom, while dipole-dipole interactions mediate ``spin"-dependent coupling between molecules. For molecules on neighboring sites, the strength of these dipole-dipole interactions is $\sim 1\t{kHz}$ for KRb and $\sim$$100\t{ kHz}$ for LiCs. These interactions are much stronger than the exchange interactions between ultracold atoms ($\ll 1\t{ kHz}$ \cite{Trotzky2008}). Therefore, polar molecules seem to be better candidates for the simulation of certain condensed matter phenomena.

Recently, it was shown that polar molecules in optical lattices can be used to simulate a highly tunable generalization of the $t$-$J$ Hamiltonian, referred to as the $t$-$J$-$V$-$W$ Hamiltonian \cite{Gorshkov2011,Gorshkov2011b}. The latter differs from the $t$-$J$ Hamiltonian in the following aspects: it has anisotropic $XXZ$ spin-spin interactions $J_\perp$ and $J_z$, an independent density-density interaction $V$, a spin-density interaction $W$, and the interactions are long-range dipolar rather than nearest-neighbor. This Hamiltonian is highly tunable, and the strengths of these interactions can, in principle, be varied independently in experiments. In particular, the regime $J>t$ can be achieved, which is not possible with cold atoms where $J$ originates from second-order virtual hopping. Furthermore, in Ref.~\cite{Gorshkov2011b}, it was shown that, in one-dimension, the $t$-$J$-$V$-$W$ model can support enhanced superfluidity relative to the standard $t$-$J$ model.

In this paper, we use the tunability of the $t$-$J$-$V$-$W$ Hamiltonian to find parameter regimes supporting robust $d$-wave superfluidity in one- and two-dimensional systems of weakly coupled plaquettes \cite{Yao2007,Tsai2008,Tsai2006,Karakonstantakis2011,Rey2009,Isaev2010}. Furthermore, we demonstrate that this solvable limit and the associated $d$-wave superfluidity are experimentally realizable. Finally, we believe that this limit can provide qualitative guidance for the case of the homogeneous two-dimensional lattice.

Throughout the paper, we consider, for simplicity, an average filling of three molecules per plaquette. In the simplest experimentally realizable case where $J_z = V = W = 0$, referred to as the $t$-$J_\perp$ Hamiltonian, we find three phases: a $d$-wave superfluid of bound holes, a checkerboard solid of alternating plaquettes of bound holes, and phase separation of bound holes. We find that the addition of a density-density interaction proportional to $V$ destroys the superfluid phase in the perturbative limit. We also find that an Ising spin-spin interaction proportional to $J_z$ or a spin-density interaction proportional to $W$ can enhance the superfluid phase for certain parameter regimes.

The remainder of the paper is organized as follows. In Sec.~\ref{sec:t-Jperp-Jz-V-W_ham}, we introduce the $t$-$J$-$V$-$W$ Hamiltonian and briefly discuss its experimental implementation with polar molecules in optical lattices. In Sec.~\ref{sec:t-Jperp}, we analyze the $t$-$J_\perp$ Hamiltonian in detail solving the single-plaquette Hamiltonian exactly and calculating the phase diagram perturbatively. In Sec.~\ref{sec:t-J-V-W}, we analyze the effects of the $J_z$, $V$, and $W$ terms on the superfluid phase. In Sec.~\ref{sec:experimental_realizations}, we present proposals for experimentally realizing the perturbative calculations for one and two-dimensional systems of plaquettes. Finally, in Sec.~\ref{sec:conclusions}, we present conclusions. Appendix~\ref{sec:group_theory} gives a brief summary of the group theoretic techniques used to solve the single-plaquette Hamiltonians exactly and describes the symmetries of the solutions. Appendix~\ref{sec:rep_basis} gives explicit expressions for the basis vectors of the irreducible representations discussed in Appendix~\ref{sec:group_theory}, and Appendix~\ref{sec:ground_states} gives explicit expressions for important ground-state energies and states.

\section{The $t$-$J$-$V$-$W$ Hamiltonian} \label{sec:t-Jperp-Jz-V-W_ham}

\begin{figure}
\bc \includegraphics[scale=.7]{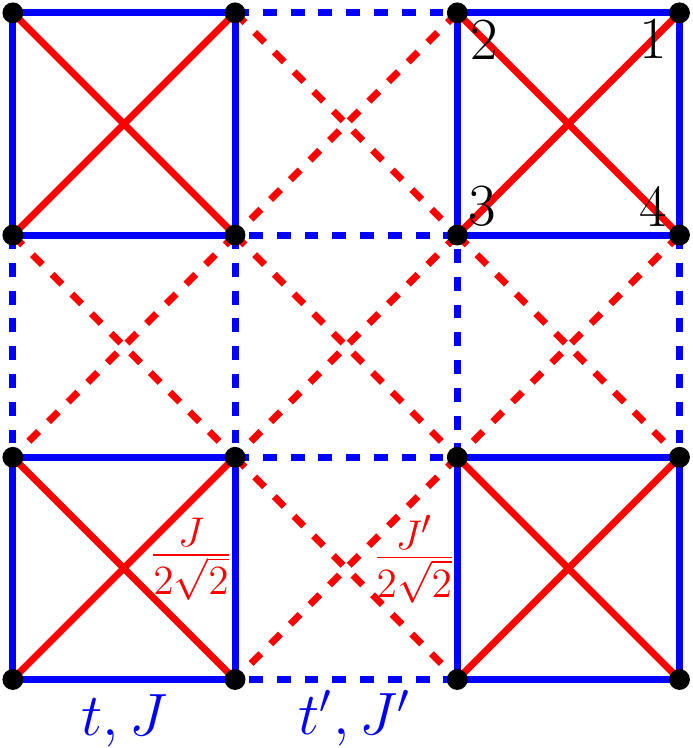} \ec
\caption{(Color online) Geometry of plaquettes. $J$ schematically denotes all dipolar interactions $J_\perp$, $J_z$, $V$, and $W$. Within each plaquette, solid blue lines contain tunneling of amplitude $t$ and dipolar interactions of strength $J$, while solid red lines only contain dipolar interactions of strength $J/(2 \sqrt{2})$. Nearest-neighbor plaquettes are then linked with tunneling of amplitude $t'$ and dipolar interactions of strength $J'$, shown as dotted blue lines, where $t'\ll t$ and $J'\ll J$. Furthermore, nearest-neighbor  and next-nearest-neighbor plaquettes are linked with dipolar interactions of strength $J'/(2 \sqrt{2})$, shown as dotted red lines. The numbering of sites in a single plaquette is shown in black.}
\label{plaquette-geometry}
\end{figure}

In the following calculations, we consider nearest-neighbor and next-nearest-neighbor interactions on a square lattice as shown in \fref{plaquette-geometry} and approximate the Hamiltonian \cite{Gorshkov2011,Gorshkov2011b} as \begin{multline} H= -\sum_{\ev{r,r'},\sigma}t_{rr'}\lp \dagg{c}_{r\sigma} c_{r'\sigma} + \dagg{c}_{r'\sigma} c_{r\sigma} \rp\\ + \lp\sum_{\ev{r,r'}} + \frac{1}{2\sqrt{2}}\sum_{\ev{\ev{r,r'}}}\rp \Bigg[ \frac{J_{\perp rr'}}{2}\lp S_r^+S_{r'}^- + S_r^-S_{r'}^+\rp\\
+J_{zrr'}S_r^zS_{r'}^z +V_{rr'}n_rn_{r'} +W_{rr'}\lp n_rS_{r'}^z +n_{r'}S_r^z\rp \Bigg],
\label{hamiltonian} \end{multline} where $\dagg{c}_{r\sigma}$ is the creation operator for a hardcore fermionic molecule at the lattice site $r$ with effective spin $\sigma$, $n_{r\sigma}=\dagg{c}_{r\sigma}c_{r\sigma}$, $n_r=n_{r\up}+n_{r\dn}$, $S_r^{+}=\dagg{c}_{r\up}c_{r\dn}$, $S_r^-=\dagg{\lp S_r^+\rp}$, and $S_r^z=\lp n_{r\up}-n_{r\dn}\rp/2$. The tunneling amplitude $t_{rr'}$ and the dipolar interaction strengths $J_{\perp rr'}$, $J_{zrr'}$, $V_{rr'}$, and $W_{rr'}$ are $t$, $J_\perp$, $J_z$, $V$, and $W$ respectively if the sites $r$ and $r'$ are in the same plaquette and are $t'$, $J'_\perp$, $J'_z$, $V'$, and $W'$ respectively if the sites are in neighboring plaquettes. The $\ev{ }$ signify that the sums are taken over nearest-neighbor bonds and the $\ev{\ev{ }}$ signify that the sums are taken over next-nearest-neighbor (diagonal) bonds. The next-nearest-neighbor bonds have a factor of $1/(2\sqrt{2})$ since they are a factor of $\sqrt{2}$ longer than the nearest-neighbor bonds and since the dipole-dipole interaction strength falls off inversely as distance cubed. \eref{hamiltonian} omits the energies of the states $\ket{\up}$ and $\ket{\dn}$ since we work at fixed numbers of up and down molecules.

The Hamiltonian given by \eref{hamiltonian} could be experimentally realized by loading ultracold polar molecules into an optical lattice and applying an external dc electric field perpendicular to the plane of the lattice \cite{Gorshkov2011,Gorshkov2011b}. Two rotational states $\ket{m_0}$ and $\ket{m_1}$ of a molecule form the effective spin states $\ket{\up}$ and $\ket{\dn}$, respectively. The preparation of these states is discussed below in Sec.~\ref{sec:preparation_detection}. Due to the dc electric field, these states have permanent electric dipole moments. The $J_z$, $V$, and $W$ interaction terms in \eref{hamiltonian} can be understood as the classical dipole-dipole interactions between these permanent dipole moments. The $J_\perp$ interaction term arises due to the transition dipole moment between $\ket{m_0}$ and $\ket{m_1}$. Large chemical reaction rates \cite{Ospelkaus2010,Idziaszek2010,Micheli2010} and large interactions between molecules on the same lattice site enforce the hardcore constraint \cite{Gorshkov2011}.

The amplitudes and signs of $J_\perp$, $J_z$, $V$, and $W$ can be tuned independently \cite{Gorshkov2011} by tuning the external dc electric field and applying external microwave fields \cite{Micheli2006,Brennen2007,Buchler2007,Wall2009,Yu2009,Krems2009,Wall2010,Schachenmayer2010,Kestner2011,Buchler2007b,Micheli2007,Gorshkov2008,Lin2010,Cooper2009}. The tunneling amplitude $t$ (assumed to be positive throughout the paper) can be tuned by adjusting the depth of the optical lattice. 

\section{Analysis of the $t$-$J_\perp$ Hamiltonian} \label{sec:t-Jperp}

The simplest experimental realization of \eref{hamiltonian} can be obtained by applying a very weak external dc electric field. In this case, the permanent electric dipole moments are very small making $J_z,$ $V,$ and $W$ negligible relative to $J_\perp$, which is proportional to the square of the transition dipole moment. In this section, we study the resulting $t$-$J_\perp$ Hamiltonian given by \eref{hamiltonian} with $J_z=V=W=0$ and $J'_z=V'=W'=0$. In Sec.~\ref{sec:t-Jperp:single_plaquette}, we describe the exact diagonalization of the $t$-$J_\perp$ Hamiltonian for a single plaquette and identify a set of conditions necessary, within our perturbative analysis, for the observation 
of $d$-wave superfluidity. In Sec.~\ref{sec:t-Jperp:double_plaquette}, we calculate the phase diagram for a two-dimensional lattice of plaquettes perturbatively. 

Throughout the remainder of this paper, we use the following notation for states. A ket with one number $\ket{n}$ refers to a single plaquette with $n=n_\up+n_\dn$ total molecules. A ket with two numbers separated by a comma $\ket{n_\up,n_\dn}$ refers to a single plaquette with $n_\up$ spin up molecules and $n_\dn$ spin down molecules. A tensor product of two kets $\ket{n_R}\ket{n_{R'}}$ refers to a system of two plaquettes with $n_R$ total molecules on plaquette $R$ and $n_{R'}$ total molecules on plaquette $R'$.

\begin{table}
\bc
\begin{tabular}{|r|r|c|}
\hline
Representation & \multicolumn{2}{|c|}{Symmetries}\\
\hline
$A_1$ & $s$ & \includegraphics[scale=.1]{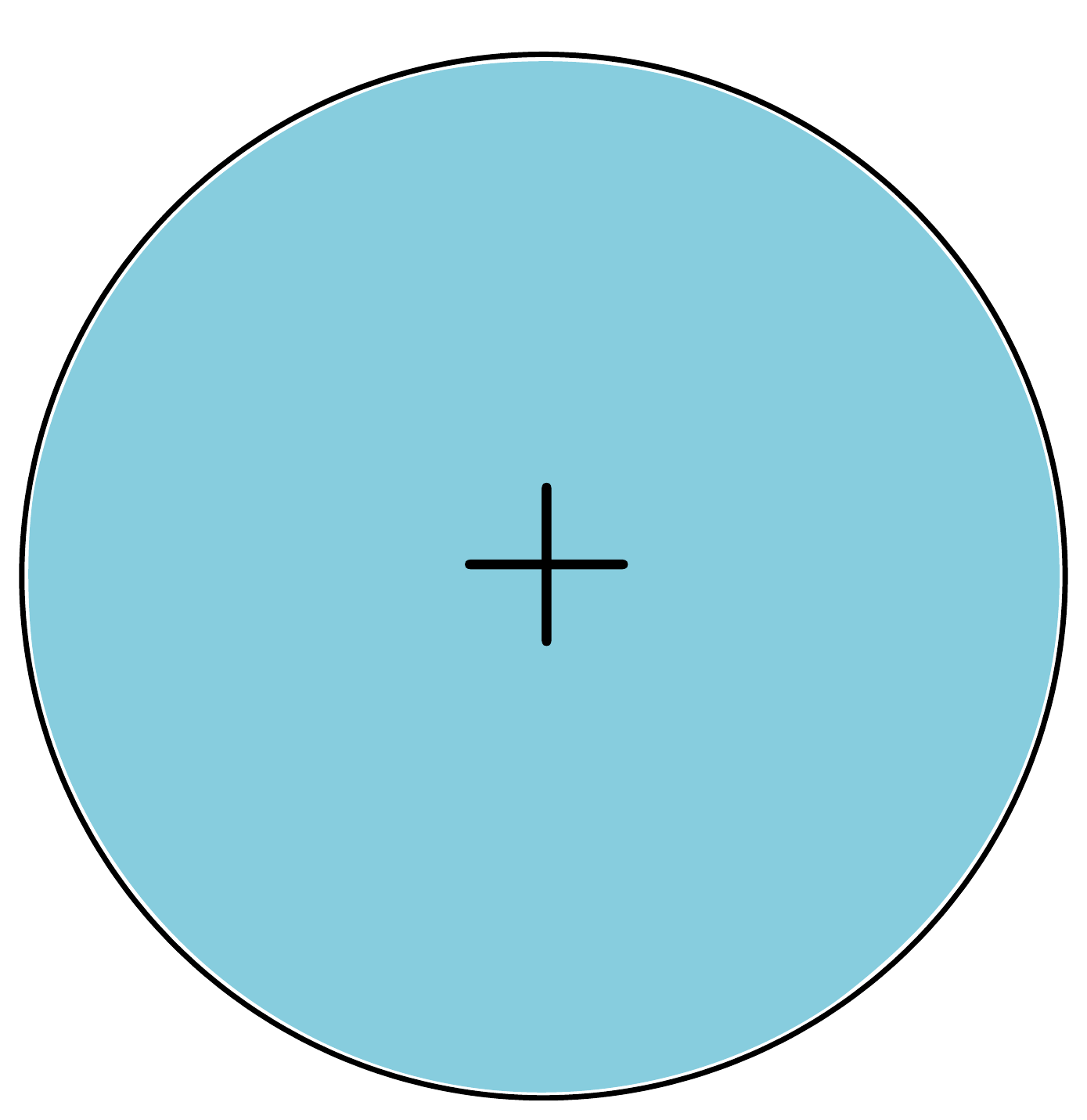}\\
\hline
$A_2$ & $s$ & \includegraphics[scale=.1]{s.pdf}\\
\hline
$B_1$ & $d_{x^2-y^2}$ & \includegraphics[scale=.1]{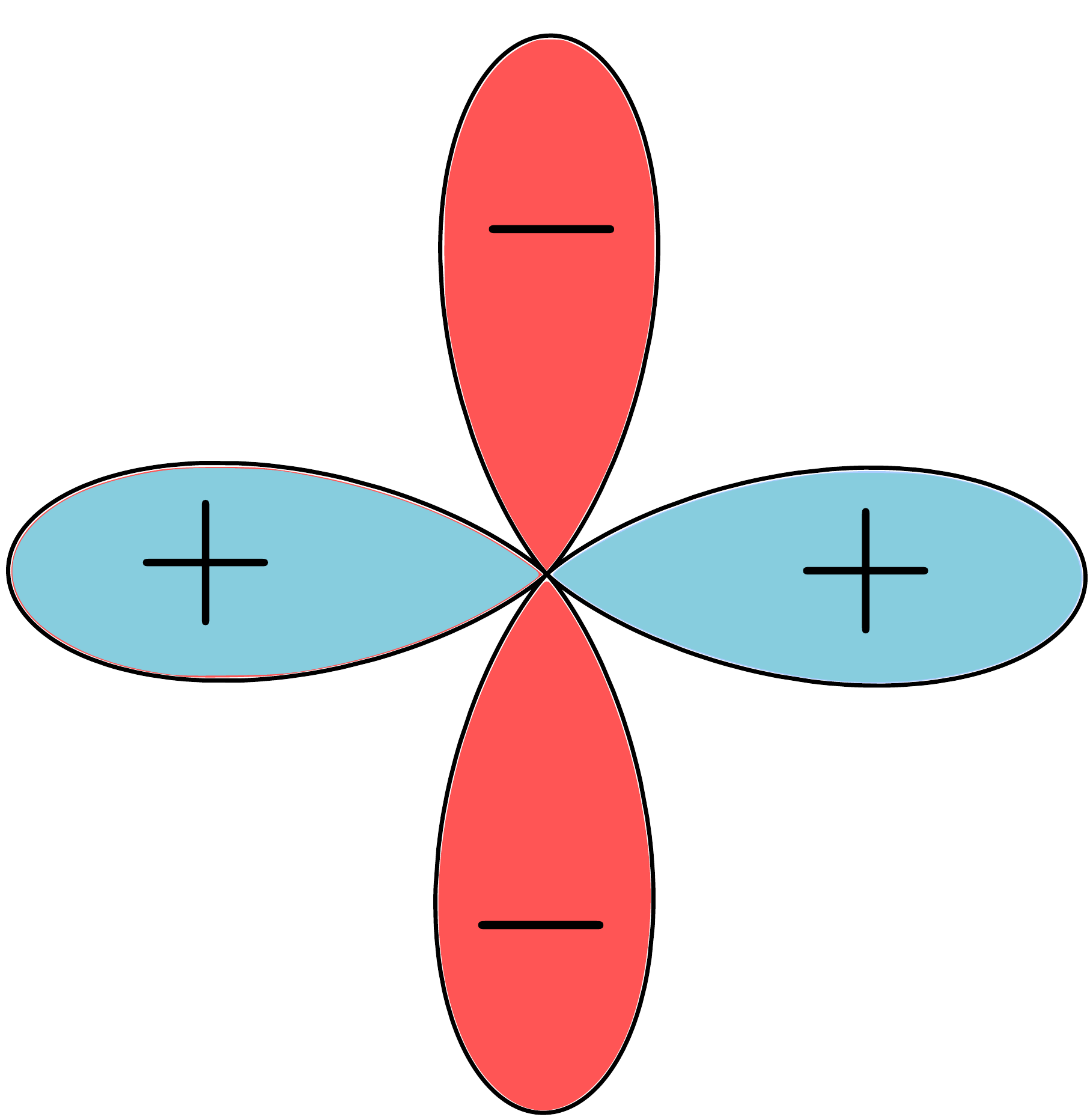}\\
\hline
$B_2$ & $d_{xy}$ & \includegraphics[scale=.1]{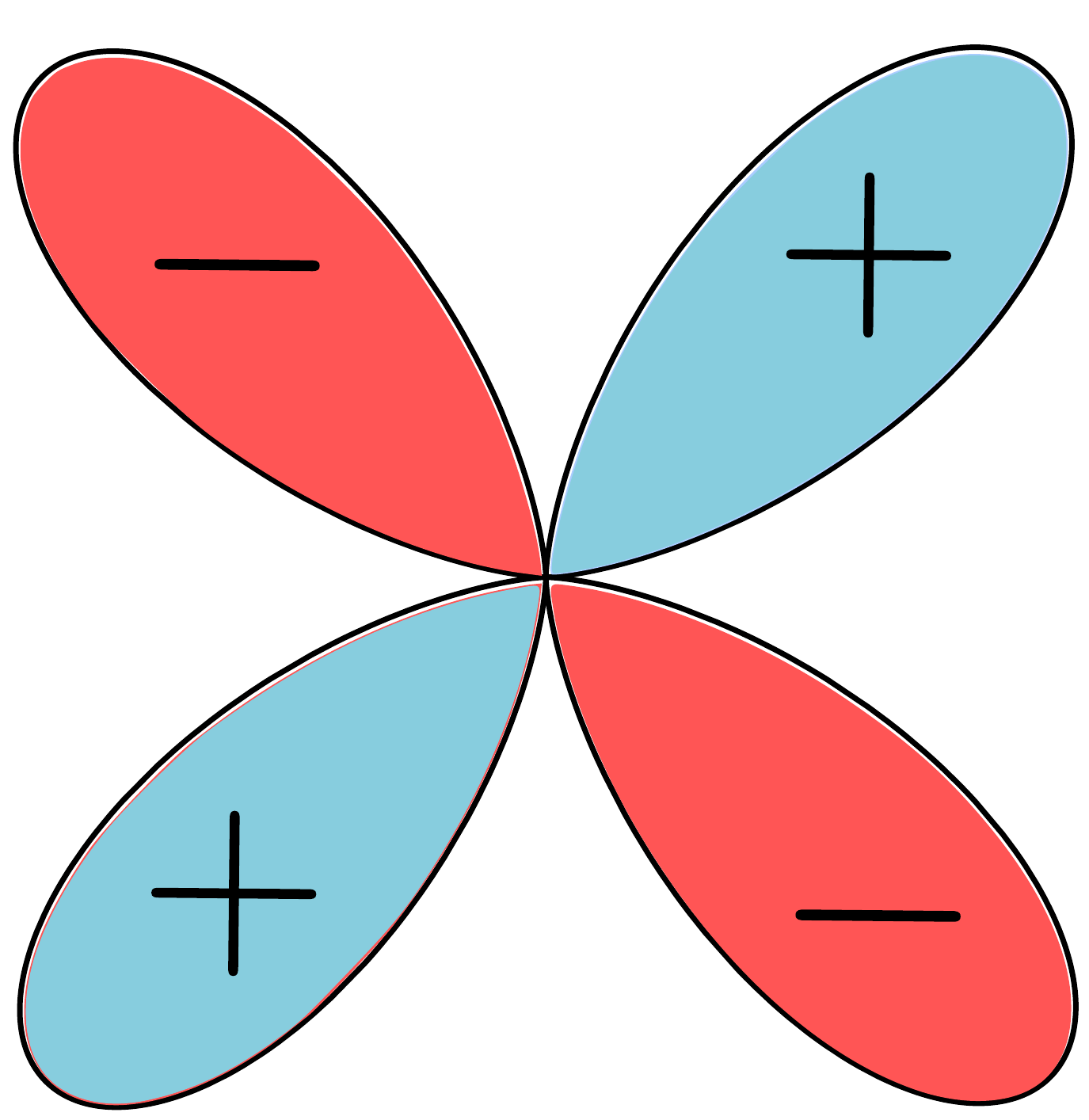}\\
\hline
$E$ & $p_x$ and $p_y$ & \includegraphics[scale=.1]{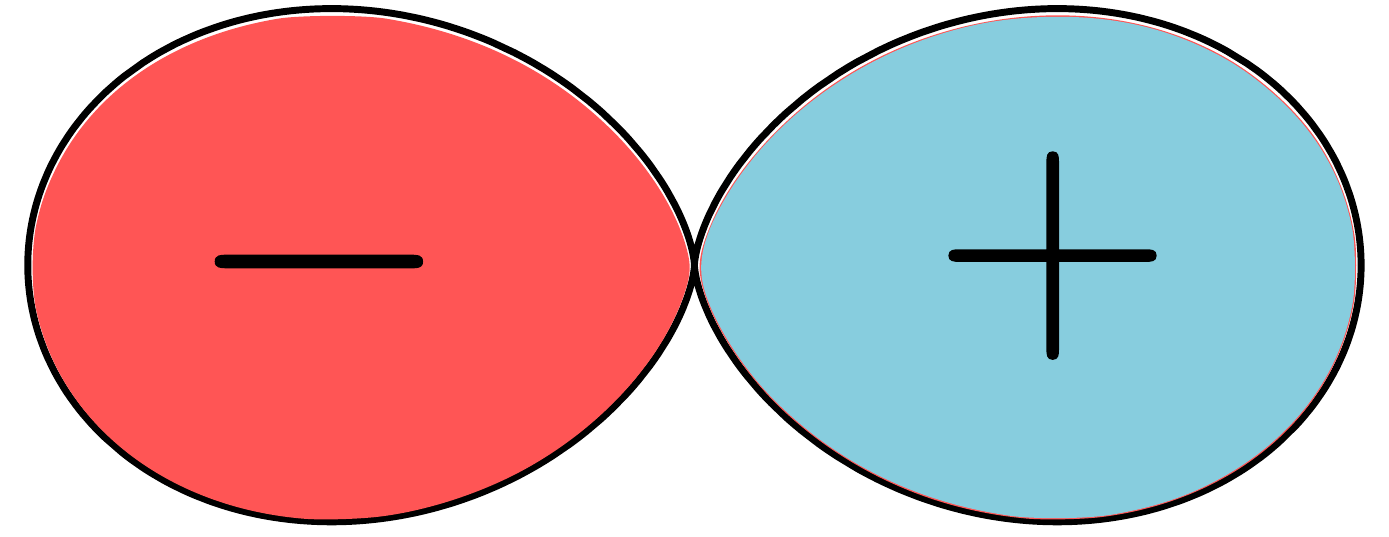} and \includegraphics[scale=.1]{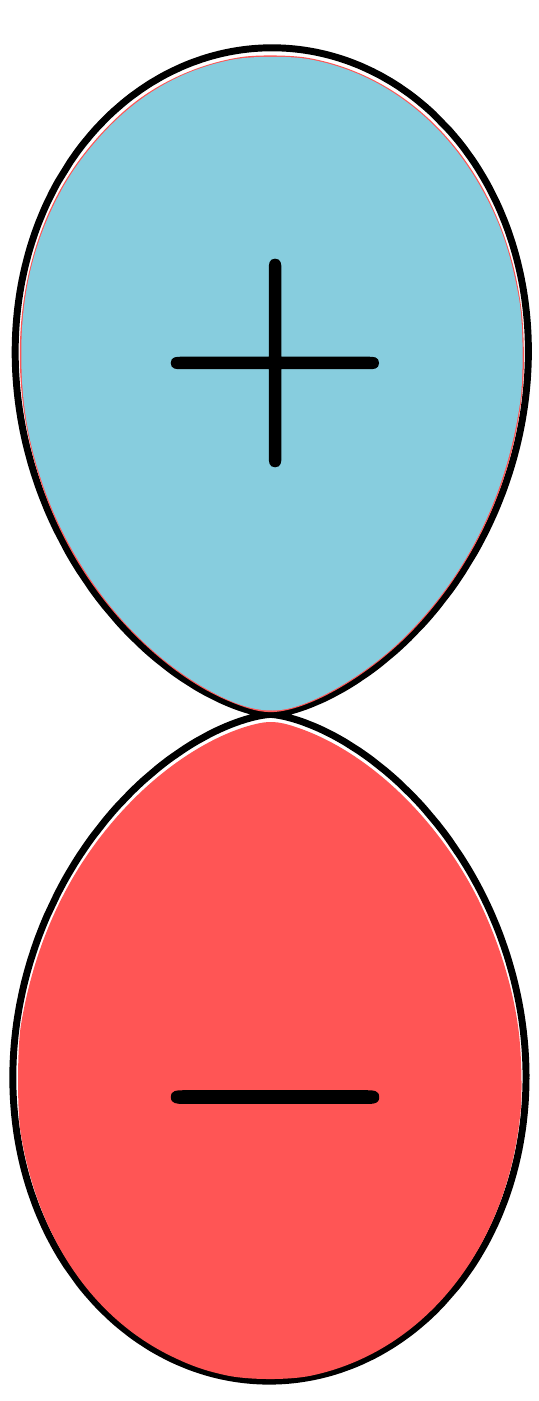}\\
\hline
\end{tabular}
\ec
\caption{(Color online) Symmetries corresponding to the irreducible representations of the group $D_4$ as is discussed in Appendix~\ref{representations-description} \cite{Tinkham1964,Yao2007,Isaev2010}. Wave function symmetries are plotted in the $xy$ plane. Note that the $A_1$ representation is symmetric under all symmetry operations of the square and that the $A_2$ representation is antisymmetric under $\pi$ rotations about the $x$ and $y$ axes and the lines $y=x$ and $y=-x$. The $A_2$ wave functions can be thought of as being positive (blue) on the front and negative (red) on the back, while all other wave functions have the same polarity on both sides of the wave function.} \label{symmetry-table}
\end{table}

\subsection{Single-Plaquette Analysis} \label{sec:t-Jperp:single_plaquette}

We use the point symmetries of the square, described by the group $D_4$, and the conservation laws of the Hamiltonian to simplify the task of diagonalizing the single-plaquette Hamiltonian with Hilbert space dimension $3^4=81$ and to understand the symmetries of the resulting eigenstates. First, the operators $n_{\up}$ and $n_{\dn}$ (which measure the number of up and down molecules on a single plaquette) commute with the Hamiltonian and with each other so we can diagonalize subspaces with fixed values of $n_{\up}$ and $n_{\dn}$ separately. This reduces the subspace dimensions to 12 for the largest subspaces. We then use the basis functions of the group $D_4$ to diagonalize the Hamiltonian for each irreducible representation separately. This requires diagonalizing $3\cross3$ matrices at worst. The symmetries of the resulting eigenstates correspond to the symmetries of the irreducible representations summarized in \tref{symmetry-table}.

\begin{figure}
\bc \includegraphics[scale=1]{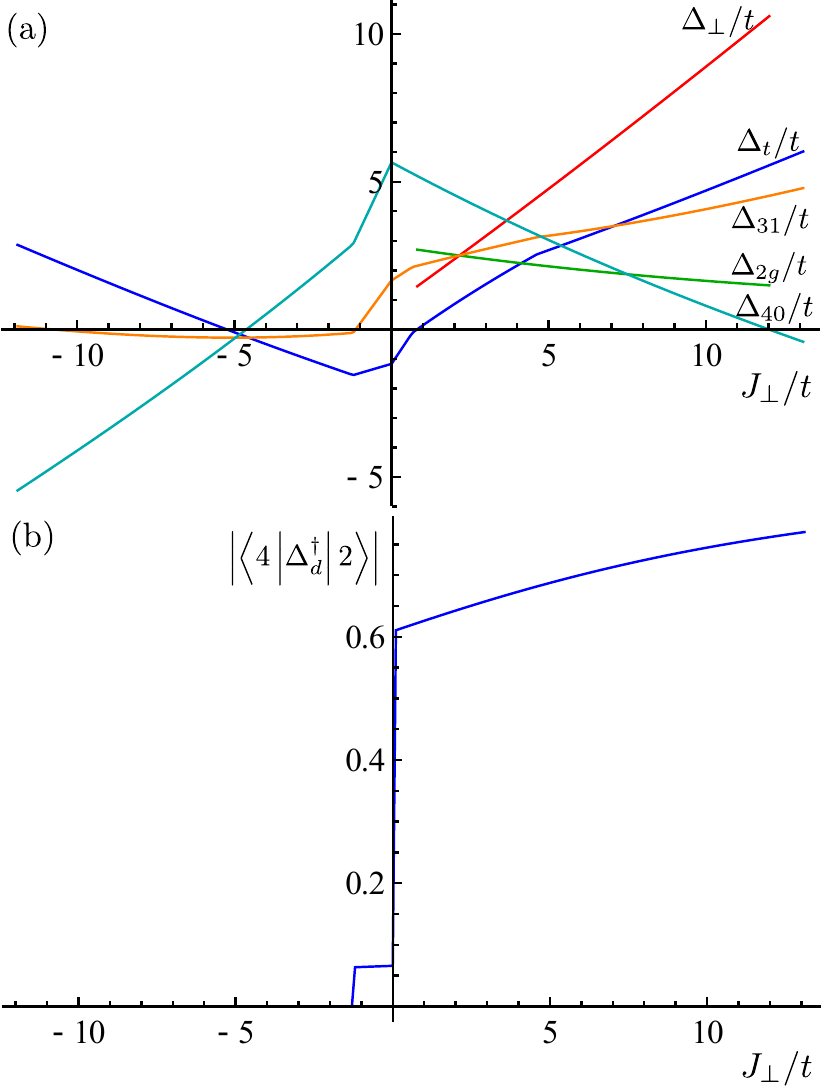} \ec \caption{(Color online) (a): Binding energy $\Delta_t$ for two holes and other relevant energies $\Delta_\perp$, $\Delta_{31}$,  $\Delta_{40}$, and $\Delta_{2g}$, all of which must be positive for the Hamiltonian to support $d$-wave superfluidity within our treatement. (b): $\abs{\Braket{4}{\dagg{\Delta}_d}{2}}$ for a single plaquette, which must be nonzero for the Hamiltonian to support $d$-wave superfluidity within our treatment.} \label{t-Jperp_all}
\end{figure}

The Hamiltonians for subspaces of constant $n_\up$ and $n_\dn$ were diagonalized in the representation basis using the methods described below in Appendix~\ref{sec:group_theory:diagonalization}. The ground states of a single plaquette with fixed total number of molecules $n=n_\up+n_\dn$ are summarized in \tref{groundstates}. In order to construct robust $d$-wave superfluids at $3/4$ filling (three molecules per plaquette), we would like it to be energetically favorable for two holes to condense on the same plaquette. This condition is achieved if the binding energy of two holes \cite{Yao2007,Tsai2008,Tsai2006,Karakonstantakis2011,Rey2009,Isaev2010} \beq \Delta_t=2E_g(3)-E_g(4)-E_g(2) \eeq is positive, where $E_g(n)$ is the energy of the ground state of $\ket{n}$. $\Delta_t$ is shown in \fref{t-Jperp_all}(a).  The $d$-wave matrix element between $\ket{2}$ and $\ket{4}$ $$\Braket{4}{\dagg{\Delta}_d}{2}$$ must also be nonzero for $d$-wave superfluidity to be possible. $\dagg{\Delta}_d$ is the $d$-wave symmetric pair creation operator defined as \cite{Rey2009,Trebst2006} \beq \dagg{\Delta}_d= \frac{1}{2}\lp \dagg{s}_{12} + \dagg{s}_{34} -\dagg{s}_{14} -\dagg{s}_{23}\rp,\eeq where \beq \dagg{s}_{rr'}=\frac{1}{\sqrt{2}} \lp\dagg{c}_{r\up}\dagg{c}_{r'\dn} -\dagg{c}_{r\dn}\dagg{c}_{r'\up}\rp \label{singlet-def} \eeq creates a singlet between sites $r$ and $r'$. Since $\dagg{s}_{rr'}=\dagg{s}_{r'r}$, $\dagg{\Delta}_d$ has $d_{x^2-y^2}$ symmetry and creates a pair of molecules with $d_{x^2-y^2}$ symmetry. Thus the condition $\Braket{4}{\dagg{\Delta}_d}{2}\neq 0$ ensures $d$-wave symmetry of bound hole pairs. $\abs{\Braket{4}{\dagg{\Delta}_d}{2}}$ is shown in \fref{t-Jperp_all}(b).

From the information in \tref{groundstates}, we can understand the behavior of $\Delta_t$ and $\abs{\Braket{4}{\dagg{\Delta}_d}{2}}$ shown in \fref{t-Jperp_all}. We see that for all values of $J_\perp/t$, the ground state of $\ket{4}$ is a $\ket{2,2}$ in the $B_1$ representation and thus always has $d_{x^2-y^2}$ symmetry. For $J_\perp/t<-1.22$, the ground state of $\ket{2}$ is a $\ket{1,1}$ in the $E$ representation and thus cannot exhibit $d$-wave superfluidity. For $J_\perp/t>-1.22$ the ground state changes to a $\ket{1,1}$ in the $A_1$ representation and thus has $s$-wave symmetry. This change in representations accounts for $\abs{\Braket{4}{\dagg{\Delta}_d}{2}}$ becoming nonzero at $J_\perp/t=-1.22$ and the kink in $\Delta_t$ at the same value. The change in the $\ket{4}$ ground state at $J_\perp/t=0$, while retaining the $d_{x^2-y^2}$ symmetry, accounts for the large jump in $\abs{\Braket{4}{\dagg{\Delta}_d}{2}}$ at this value.

For $0<J_\perp/t<0.66$, the ground states of $\ket{3}$ are a $\ket{3,0}$ and a $\ket{0,3}$ in the $A_2$ representation; for all other values of $J_\perp/t$, the ground states are a $\ket{2,1}$ and a $\ket{1,2}$. At $J_\perp/t=0.66$ the state changes to the $E$ representation and at $J_\perp/t=4.62$ the state changes to the $B_1$ representation. These changes in representations account for the kinks in $\Delta_t$ at these values of $J_\perp/t$. $\Delta_t$ crosses the positive $J_\perp/t$ axis at $J_\perp/t=0.82$. For $J_\perp/t>0.82$, the condition $\abs{\Braket{4}{\dagg{\Delta}_d}{2}} \neq 0$ also holds. Thus in this regime, holes bind into $d$-wave symmetric pairs, which are necessary for $d$-wave superfluidity.

The $t$-$J_\perp$ Hamiltonian is an improvement over the Hubbard Hamiltonian in that the binding energy $\Delta_t$ for the Hubbard Hamiltonian is positive only for a narrow parameter region $0<U/t<4.6$ \cite{Rey2009}. Furthermore, the binding energy reaches a maximum for the Hubbard Hamiltonian \cite{Rey2009} while it increases with $J_\perp/t$ for the $t$-$J_\perp$ Hamiltonian.

\begin{table}
\bc
\begin{tabular}{|r|r|r|}
\hline
$J_\perp/t$ & $n_\up,n_\dn$ & Representation \\
\hline
$\in (-\infty,\infty)$ & 1,0 & $A_1$ \\
\hline
$<-1.22$ & 1,1 & $E$ \\
$>-1.22$ & 1,1 & $A_1$ \\
\hline
$<0$ & 2,1 & $A_2$ \\
$\in(0,0.66)$ & 3,0 & $A_2$ \\
$\in(0.66,4.62)$ & 2,1 & $E$ \\
$>4.62$ & 2,1 & $B_1$ \\
\hline
$<0$ & 2,2 & $B_1$ \\
$>0$ & 2,2 & $B_1$ \\
\hline

\end{tabular}
\ec
\caption{Ground-state configurations and symmetries of states with constant $n=n_\up+n_\dn$. The results are symmetric under interchange of up spins with down spins. The ground state for four molecules is always $\ket{2,2}$ in the $B_1$ representation; however, there are two copies of the $B_1$ representation for the $\ket{2,2}$ subspace, see Appendix~\ref{sec:group_theory}, and the particular ground state changes at $J_\perp/t=0$ even though the symmetry of the ground state remains $d_{x^2-y^2}$.}\label{groundstates}
\end{table}

For positive values, the binding energy $\Delta_t$ is the amount that the ground state $\ket{2}\ket{4}$ or $\ket{4}\ket{2}$ is energetically favorable over the ground state $\ket{3}\ket{3}$, which could be coupled to $\ket{2}\ket{4}$ and $\ket{4}\ket{2}$ through tunneling $t'$. For positive values of $\Delta_t$ it is also necessary to consider the energy difference between the lowest energy states coupled to $\ket{2}\ket{4}$ and $\ket{4}\ket{2}$ through the spin interaction $J'_\perp$: $$\Delta_\perp=E_g(0,2)+E_g(3,1) - E_g(1,1)- E_g(2,2),$$ where $E_g(n_\up,n_\dn)$ is the energy of the lowest $\ket{n_\up,n_\dn}$ state. $\Delta_\perp$ is shown in \fref{t-Jperp_all}(a) for values of $J_\perp/t$ where $\Delta_t>0$ and $\abs{\Braket{4}{\dagg{\Delta}_d}{2}}\neq0$. The size of the binding energies $\Delta_t$ and $\Delta_\perp$ roughly correspond to how large $t'$ and $J'_\perp$ can be, respectively.

For $J_\perp/t>-1.22$, the ground state of $\ket{2}$ is an $s$-wave $\ket{1,1}$ in the $A_1$ representation. However, another $\ket{1,1}$ state in the $E$ representation becomes close in energy to the ground state for large $J_\perp/t$. So we define a third energy difference $$\Delta_{2g}=E_g(1,1(E))-E_g(1,1(A_1))$$ to quantify the energy gap between these two $\ket{2}$ states. Here $E_g(1,1(\Gamma))$ is the energy of the lowest $\ket{1,1}$ state in the irreducible representation $\Gamma$. $\Delta_{2g}$ roughly corresponds to how large the overall perturbing Hamiltonian $H_{\t{eff}}$ linking plaquettes can be. $\Delta_{2g}$ is also shown in \fref{t-Jperp_all}(a) for values of $J_\perp/t$ where $\Delta_t>0$ and $\Braket{4}{\dagg{\Delta}_d}{2}\neq0$.

Provided that $\Delta_t>0$, in a full lattice of plaquettes with $3/8$ of the lattice sites occupied by up molecules and $3/8$ of the lattice sites occupied by down molecules, it could, in principle, be energetically favorable for the molecules to arrange themselves in configurations other than two and four molecules on a plaquette throughout the entire lattice. Since $\Delta_t>0$, it will be energetically costly to change $\ket{2}\ket{4}$ to $\ket{3}\ket{3}$. Two four molecule plaquettes $\ket{4}\ket{4}$ cannot rearrange their molecules since they are at maximum filling. However, it is possible for two two molecule plaquettes $\ket{2}\ket{2}$ to rearrange themselves to $\ket{3}\ket{1}$ or $\ket{4}\ket{0}$. Thus it is also necessary to consider the binding energies $$\Delta_{31}=E_g(3)+E_g(1)-2E_g(2)$$ and $$\Delta_{40}=E_g(4)+E_g(0)-2E_g(2)=E_g(4)-2E_g(2).$$ $\Delta_{31}$ and $\Delta_{40}$ are shown in \fref{t-Jperp_all}(a).

As is shown in \fref{t-Jperp_all}(a), when $\Delta_t$ is positive, the other relevant energies are also positive for $J_\perp/t$ less than about 12. For $J_\perp/t$ greater than about 12, $\Delta_{40}$ becomes negative, and it becomes energetically favorable for a $\ket{2}\ket{2}$ to change to a $\ket{4}\ket{0}$ or $\ket{0}\ket{4}$, in which case the manifold of states consisting of only $\ket{2}$ and $\ket{4}$ plaquettes stops being the true ground state. However, the corresponding phase diagram can still be studied experimentally by adiabatically preparing these -- no longer ground -- states beginning with easily preparable excited states. A calculation to minimize the energy of a full lattice of plaquettes with $t'=J'_\perp=0$ confirms that a lattice with half of the plaquettes as $\ket{2}$ and half as $\ket{4}$ is the ground state when $\Delta_{40}$ and $\Delta_t$ are positive.

Although it is outside the scope of the present paper, we point out that the narrow region $0<J_\perp/t<0.66$, where $\ket{0,3}$ and $\ket{3,0}$ are the degenerate ground states, could support Nagaoka ferromagnetism \cite{VonStecher2010,Nagaoka1966}.

\subsection{Double-Plaquette Analysis} \label{sec:t-Jperp:double_plaquette}

In this section, we describe the behavior of the full lattice of plaquettes. We use Schrieffer-Wolff transformations to find the interactions between nearest-neighbor and next-nearest-neighbor plaquettes to second order in $t'/t$ and $J'_\perp/t$. We then derive an effective $XXZ$ Hamiltonian and solve for the phase diagram in the perturbative limit.

\subsubsection{Schrieffer-Wolff Transformations} \label{sec:schrieffer-wolff}

We solve the problem of two coupled plaquettes with second-order perturbation theory through the Schrieffer-Wolff transformation \cite{Schrieffer1966}. There are six relevant double-plaquette problems in describing a full two-dimensional lattice. A $\ket{4}$ can interact with another $\ket{4}$ or with a $\ket{2}$, and a $\ket{2}$ can also interact with another $\ket{2}$. For each of these three cases, the plaquettes can be situated either next to each other or diagonally.

In all cases, the unperturbed Hamiltonian is the tensor product of two single-plaquette Hamiltonians 
\begin{equation}
H_0=\bbmat 0 & 0\\
0 & U \ebmat \label{eq:h0}
\end{equation}
written here in the basis that diagonalizes it. If the dimension of the two-plaquette Hilbert space is $d$ and the ground state is $l$-fold degenerate, then the upper left $0$ is an $l\cross l$ zero matrix and $U$ is a $(d-l)\cross(d-l)$ diagonal matrix with the energy differences between the ground and excited states along the diagonal. The $\ket{2}$ and $\ket{4}$ energies are different; however, since the number of $\ket{2}$ and $\ket{4}$ plaquettes is constant, we drop these energies here.

The perturbing Hamiltonian is $$H_1= \bbmat H_{1g} & H_{1ge}\\
H^T_{1ge} & H_{1e} \ebmat,$$ where $H_{1g}$ defines the first-order shifts in the energies of the ground states due to the perturbation, $H_{1ge}$ defines the couplings between the ground and excited states, and $H_{1e}$ defines the couplings between the excited states and the first-order shifts in the energies of the excited states. The effective Hamiltonian for the low-energy subspace is then \beq H_{\t{eff}}=H_{1g}-H_{1ge}U^{-1}H_{1ge}^T+\dots. \label{effective_ham} \eeq For the remainder of the paper, we divide the Hamiltonians by $t$. Then the first term is of order $t'/t$ and $J'_\perp/t$, while the second term is of order $(t'/t)^2$ and $(J'_\perp/t)^2.$

We then have the following effective Hamiltonians when the two plaquettes are situated next to each other
\begin{subequations}
\beqnn H_{\t{eff}}^{(4,2)}\e \bbmat f^{(4,2)} & g^{(4,2)}\\
g^{(4,2)} & f^{(4,2)} \ebmat \qquad\t{ basis } \bbmat \ket{2}\ket{4}\\
\ket{4}\ket{2}\ebmat, \\
H_{\t{eff}}^{(2,2)}\e f^{(2,2)} \qquad\qquad\qquad\t{ basis } \ket{2}\ket{2},\\
H_{\t{eff}}^{(4,4)}\e f^{(4,4)} \qquad\qquad\qquad\t{ basis } \ket{4}\ket{4}, \eeqnn
\label{heff-nn}
\end{subequations}
where
\begin{subequations}
\beqnn
f^{(4,2)}\e \lp\frac{t'}{t}\rp^2 f_t^{(4,2)}\lp\jpot\rp +\lp\frac{J'_\perp}{t}\rp^2 f_\perp^{(4,2)} \lp\jpot\rp,\\
g^{(4,2)}\e \lp\frac{t'}{t}\rp^2 g_t^{(4,2)}\lp\jpot\rp,\\
f^{(2,2)}\e \lp\frac{t'}{t}\rp^2 f_t^{(2,2)}\lp\jpot\rp +\lp\frac{J'_\perp}{t}\rp^2 f_\perp^{(2,2)} \lp\jpot\rp,\\
f^{(4,4)}\e\lp\frac{J'_\perp}{t}\rp^2 f_\perp^{(4,4)}\lp\jpot\rp.
\eeqnn
\label{sw-functions-1}
\end{subequations} Here the functions $f$ and $g$ on the right hand sides depend on the interaction strength $J_\perp/t$ and describe the perturbative coupling of the plaquettes. The superscript $(n_R,n_{R'})$ refers to the number of molecules on neighboring plaquettes $R$ and $R'$. The subscripts $t$ and $\perp$ refer to interplaquette couplings driven by $t'$ and $J'_\perp$, respectively. For the $t$-$J_\perp$ Hamiltonian, there are no first-order shifts in the ground-state energies so there are no terms proportional to $J'_\perp/t$. The $\ket{2}\ket{4}$ and $\ket{4}\ket{2}$ states are only coupled through tunneling to second order, so there is no $g_\perp^{(4,2)}$ function. The $\ket{4}\ket{4}$ state cannot couple to itself through tunneling since both plaquettes are fully occupied so there is no $f_t^{(4,4)}$ function.

The effective Hamiltonians for two plaquettes situated diagonally are
\begin{subequations}
\beqnn
 H_{\t{eff}}^{(4,2)}\e \bbmat h^{(4,2)} & 0\\
0 & h^{(4,2)} \ebmat \qquad\t{ basis } \bbmat \ket{2}\ket{4}\\
\ket{4}\ket{2}\ebmat, \\
H_{\t{eff}}^{(2,2)}\e h^{(2,2)} \qquad\qquad\qquad\t{ basis } \ket{2}\ket{2},\\
H_{\t{eff}}^{(4,4)}\e h^{(4,4)} \qquad\qquad\qquad\t{ basis } \ket{4}\ket{4}, \eeqnn
\label{heff-nnn}
\end{subequations}
where
\begin{subequations}
\beqnn
h^{(4,2)}\e \lp\frac{J'_\perp}{t}\rp^2 h_\perp^{(4,2)}\lp\jpot\rp,\\
h^{(2,2)}\e \lp\frac{J'_\perp}{t}\rp^2 h_\perp^{(2,2)}\lp\jpot\rp,\\
h^{(4,4)}\e \lp\frac{J'_\perp}{t}\rp^2 h_\perp^{(4,4)}\lp\jpot\rp.
\eeqnn
\label{sw-functions-2}
\end{subequations}
When the plaquettes are situated diagonally, there is no tunneling between them, so there are no terms proportional to $(t'/t)^2$. In particular, the $\ket{2}\ket{4}$ and $\ket{4}\ket{2}$ states cannot couple to each other through tunneling in second order, so the off-diagonal terms are zero.

\subsubsection{$XXZ$ Effective Hamiltonian} \label{sec:xxz_hamiltonian}

The full lattice of plaquettes can be mapped to an $XXZ$ spin Hamiltonian \cite{Yao2007,Tsai2006,Rey2009,Isaev2010} where each plaquette becomes a lattice site, labelled by $R$, and the states $\ket{2}$ and $\ket{4}$ of two and four molecules become the effective spin up $\ket{\Up}$ and spin down $\ket{\Dn}$ states, respectively. Using the functions \eref{sw-functions-1} and \eref{sw-functions-2} calculated using the Schrieffer-Wolff transformation, the effective Hamiltonian is \begin{multline} H_{\t{eff}}= \sum_{\ev{R,R'}}\Big[ f^{(4,2)}\lp n_{R\Up}n_{R'\Dn} + n_{R\Dn}n_{R'\Up}\rp+\\
g^{(4,2)}\lp S_{R}^+S_{R'}^- + S_R^-S_{R'}^+\rp +f^{(2,2)} n_{R\Up}n_{R'\Up}
+ f^{(4,4)} n_{R\Dn}n_{R'\Dn} \Big]\\
 +\sum_{\ev{\ev{R,R'}}}\Big[ h^{(4,2)}\lp n_{R\Up}n_{R'\Dn} + n_{R\Dn}n_{R'\Up}\rp\\
 +h^{(2,2)} n_{R\Up}n_{R'\Up}+ h^{(4,4)} n_{R\Dn}n_{R'\Dn} \Big].\non\end{multline} Since each site $R$ has either one spin up or one spin down, $$n_{R\Up}=\frac{1}{2}+S_R^Z \qquad\t{ and }\qquad n_{R\Dn}=\frac{1}{2}-S_{R}^Z.$$ Thus, dropping constant terms, \begin{multline} H_{\t{eff}} =  \sum_{\ev{R,R'}} \lb \tilde{J}_\perp\lp S_R^XS_{R'}^X +S_R^YS_{R'}^Y\rp +\tilde{J}_{z1} S_R^ZS_{R'}^Z\rb\\
+ \sum_{\ev{\ev{R,R'}}} \tilde{J}_{z2} S_R^ZS_{R'}^Z + \tilde{B}\sum_R S_R^Z, \label{xxz-hamiltonian} \end{multline} where
\begin{subequations}
\beqnn \tilde{J}_\perp\e 2g^{(4,2)}, \label{effective_strengths-a}\\
\tilde{J}_{z1}\e f^{(2,2)}+ f^{(4,4)} -2f^{(4,2)}, \label{effective_strengths-b}\\
\tilde{J}_{z2}\e h^{(2,2)}+ h^{(4,4)} -2h^{(4,2)}, \label{effective_strengths-c}\\
\tilde{B}\e 2\lp f^{(2,2)} -f^{(4,4)} +h^{(2,2)}- h^{(4,4)}\rp. \label{effective_strengths-d}\eeqnn
\label{effective_strengths}
\end{subequations} Since we are interested in the phase diagram at constant $3/4$ filling of the lattice with molecules, $\sum_R S_R^Z=0$ is constant in the $XXZ$ effective Hamiltonian, so we can neglect the magnetic field term $\tilde B$. In fact, for the same reason, we have already dropped the energies of $\ket{2}$ and $\ket{4}$ in $H_0$ [\eref{eq:h0}].

\subsubsection{Phase Diagram for the $t$-$J_\perp$ Hamiltonian} \label{sec:t-Jperp:phase_diagram}

If we only consider nearest-neighbor interactions, then $\tilde{J}_{z2}=0$. In this case, there are three phases, which can be qualitatively understood by considering the following limits of the $XXZ$ model \cite{Giamarchi2003}. If $\tilde{J}_{z1}\gg |\tilde{J}_\perp|$, it is energetically favorable for the spins to be anti-ferromagnetically ordered in the $Z$ direction corresponding to a checkerboard solid of bound holes. If $\tilde{J}_{z1}\ll-|\tilde{J}_\perp|$, it is energetically favorable for the spins to be ferromagnetically ordered in the $Z$ direction; however, since $\sum_{R}S_R^Z$ is fixed, this corresponds to a phase separation of the spins and a phase separation of the bound holes. Finally, if $|\tilde{J}_\perp| \gg \abs{\tilde{J}_{z1}}$, it is energetically favorable for the spins to be  ordered in the $XY$ plane, which corresponds to a superfluid of bound holes. Specifically, for $\tilde{J}_\perp < 0$ ($\tilde{J}_\perp > 0$), the spin order is ferromagnetic (anti-ferromagnetic), corresponding to a $d$-wave superfluid with correlation function $\langle \Delta_{d,R}^\dagger \Delta_{d,R'}\rangle$ whose sign has uniform (checkerboard) structure in the $R-R'$ plane. Since $\tilde{J}_\perp$ can be mapped to $-\tilde{J}_\perp$ by a sublattice rotation  \cite{Giamarchi2003}, we do not distinguish, for $|\tilde{J}_\perp| \gg \abs{\tilde{J}_{z1}}$, between the ferromagnetic and anti-ferromagnetic cases and simply refer to both phases as a $d$-wave superfluid. The phase transitions occur at $\abs{\tilde{J}_\perp}=\abs{\tilde{J}_{z1}}$ \cite{Giamarchi2003,Hebert2001}. 

The case of nonzero next-nearest-neighbor interactions, nonzero $\tilde{J}_{z2}$, has been studied numerically in Ref.~\cite{Hebert2001} and with mean field theory in Refs.~\cite{Batrouni1995,Scalettar1995,Batrouni2000}.
For example, in Ref.~\cite{Hebert2001}, it is shown that, for a certain parameter range satisfying $\tilde{J}_{z2} \gtrsim \tilde{J}_{z1} > 0$ and $\tilde{J}_{z2} \gtrsim \tilde{J}_\perp > 0$,
it is energetically favorable for the plaquettes to arrange themselves in a striped solid.
Assuming, by analogy, that $|\tilde{J}_{z2}| \gtrsim |\tilde{J}_{z1}|$ and $|\tilde{J}_{z2}| \gtrsim |\tilde{J}_\perp|$ are both necessary for a new phase to appear, no such phase can occur in the perturbative phase diagram for the $t$-$J_\perp$ Hamiltonian since this set of conditions is never satisfied. Furthermore, near the phase transition boundaries for this phase diagram, the magnitude of $\tilde{J}_{z2}$ is about an order of magnitude smaller than the magnitude of $\tilde{J}_{z1}$. We therefore expect $\tilde{J}_{z2}$ to have an insignificant effect on the locations of phase transitions, so we neglect $\tilde{J}_{z2}$ for the remainder of the paper.

\begin{figure}
\bc \includegraphics[scale=1]{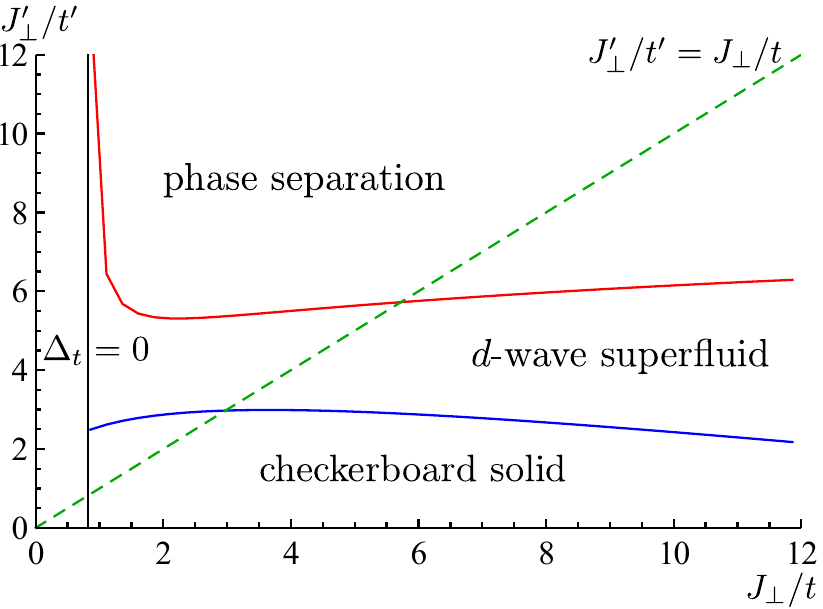} \ec
\caption{(Color online) Phase diagram for the $t$-$J_\perp$ Hamiltonian. The region to the right of the vertical black line is the region where $\Delta_t>0$. The line $J'_\perp/t'=J_\perp/t$ is shown as a dashed green line and passes through all three phases.}
\label{phase-diagram-t-Jperp}
\end{figure}

The condition for a phase transition $\pm\abs{\tilde{J}_\perp}= \tilde{J}_{z1}$ is thus \begin{multline} \pm 2\abs{g_t^{(4,2)}} = f_t^{(2,2)} - 2f_t^{(4,2)}\\
 + \lp\frac{J'_\perp}{t'}\rp^2 \lp f_\perp^{(2,2)} + f_\perp^{(4,4)} -2f_\perp^{(4,2)}\rp. \label{phasetrans-cond-t-Jperp}\end{multline} Here, a transition between superfluid and checkerboard solid occurs for $+$, a transition between superfluid and phase separation occurs for $-$, and the functions are evaluated at $J_\perp/t$. By solving \eref{phasetrans-cond-t-Jperp}, the phase diagram in the original variables $J_\perp/t$ and $J'_\perp/t'$ is computed in the region $\Delta_t>0$ and is shown in \fref{phase-diagram-t-Jperp}. All three phases are present in this phase diagram and the $d$-wave superfluid phase exists for a large range of values of $J_\perp/t$ and $J'_\perp/t'$. The easiest case to study experimentally is $t'=t$ and $J'_\perp=J_\perp$, which is outside the validity of this perturbative calculation. However, as a guess as to the physics for these values of $t'$ and $J'_\perp$, it is useful to consider the line $J'_\perp/t'=J_\perp/t$. As is shown in \fref{phase-diagram-t-Jperp}, this line passes through all three phases, indicating that all three phases might be observable in the simplest $t$-$J_\perp$ experiment with a homogeneous lattice.

There is a strong indication that the qualitative features of our results may be relevant to the non-perturbative regime where $J_\perp'/t' = J_\perp/t$. Specifically, the phase diagram along the line $J_\perp'/t' = J_\perp/t$ in \fref{phase-diagram-t-Jperp} is qualitatively similar to the phase diagram along the line of $1/4$ hole density in Fig.\ 4 of Ref.~\cite{Dagotto1993}, which numerically studies the standard $t$-$J$ model. Indeed, the order of the phases in
the two diagrams is the same provided that one identifies the region of uncondensed bound holes in Ref.~\cite{Dagotto1993} with our checkerboard solid phase and the Fermi liquid in Ref.~\cite{Dagotto1993} with our region $\Delta_t < 0$.

\section{Analysis of the Effects of $J_z$, $V$, and $W$} \label{sec:t-J-V-W}

In this section, we analyze the effects of $J_z$, $V$, and $W$ on the $d$-wave superfluid phase. First in Sec.~\ref{sec:t-J-V-W:single_plaquette}, we repeat the single-plaquette analysis of Sec.~\ref{sec:t-Jperp:single_plaquette} for various Hamiltonians with nonzero $J_z$, $V$, and $W$. Then in Sec.~\ref{sec:t-J-V-W_phase_diagrams}, we compute the phase diagrams using the methods of Sec.~\ref{sec:t-Jperp:double_plaquette} for those Hamiltonians capable of exhibiting $d$-wave superfluidity.

\subsection{Single-Plaquette Solutions} \label{sec:t-J-V-W:single_plaquette}

\begin{figure}
\bc \includegraphics[scale=.9]{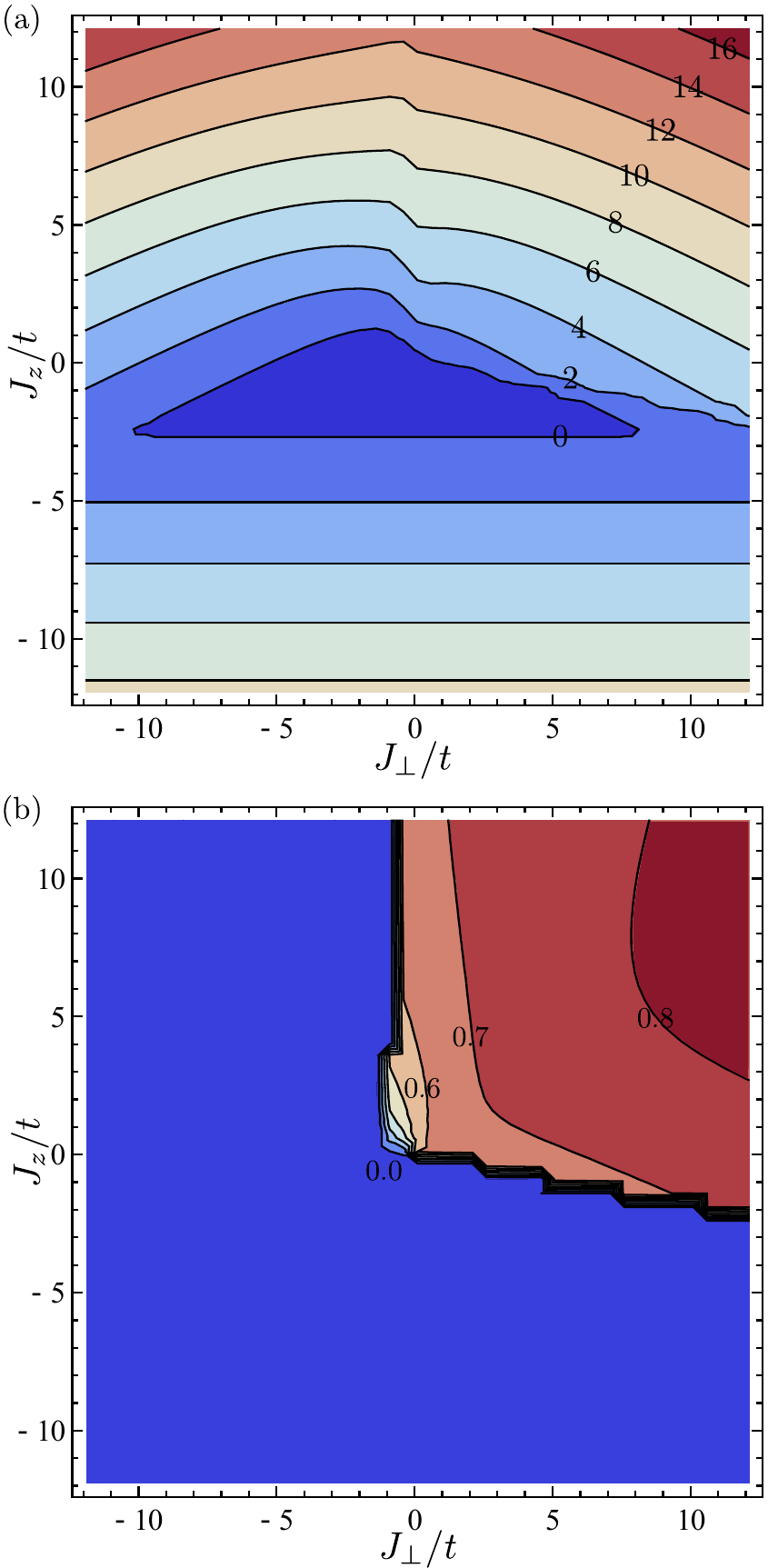} \ec
\caption{(Color online) Single-plaquette analysis for the $t$-$J_\perp$-$J_z$ Hamiltonian. (a): Binding energy $\Delta_t$. Contours of constant $\Delta_t/t$ are labeled. (b): Matrix element $\abs{\Braket{4}{\dagg{\Delta}_d}{2}}$. Contours of constant $\abs{\Braket{4}{\dagg{\Delta}_d}{2}}$ are labeled.}
\label{t-Jperp-Jz_all}
\end{figure}

First, we examine the Hamiltonian with independent $J_\perp$ and $J_z$ and zero $V$ and $W$, the $t$-$J_\perp$-$J_z$ Hamiltonian. Contour plots for the binding energy $\Delta_t$ and $\abs{\Braket{4}{\dagg{\Delta}_d}{2}}$ for this Hamiltonian are shown in \fref{t-Jperp-Jz_all}. From these figures we see that $\Delta_t>0$ and $\abs{\Braket{4}{\dagg{\Delta}_d}{2}}\neq0$ for most values of $J_\perp/t>0$ and $J_z/t>0$.

The $\ket{4}$ ground states in the $A_2$ and $B_1$ representations for positive $J_z/t$ are \beqn \ket{2,2(A_2)}\e \frac{1}{\sqrt{2}} \lp \ket{\up\dn\up\dn} - \ket{\dn\up\dn\up} \rp,\\
\ket{2,2(B_1)} \e \frac{1}{\sqrt{2}} \lp \ket{\up\dn\up\dn} + \ket{\dn\up\dn\up} \rp + \mathcal{O}\lp\frac{J_\perp}{t}\rp.
\eeqn Thus, for positive $J_z/t$ and along the line $J_\perp/t=0$, the two alternating spin configurations are the two-fold degenerate $\ket{4}$ ground states. When $J_\perp/t$ is made nonzero but small, the symmetric combination of these two states, which has $B_1$ symmetry, becomes the non-degenerate ground state.

\begin{figure}
\bc \includegraphics[scale=.9]{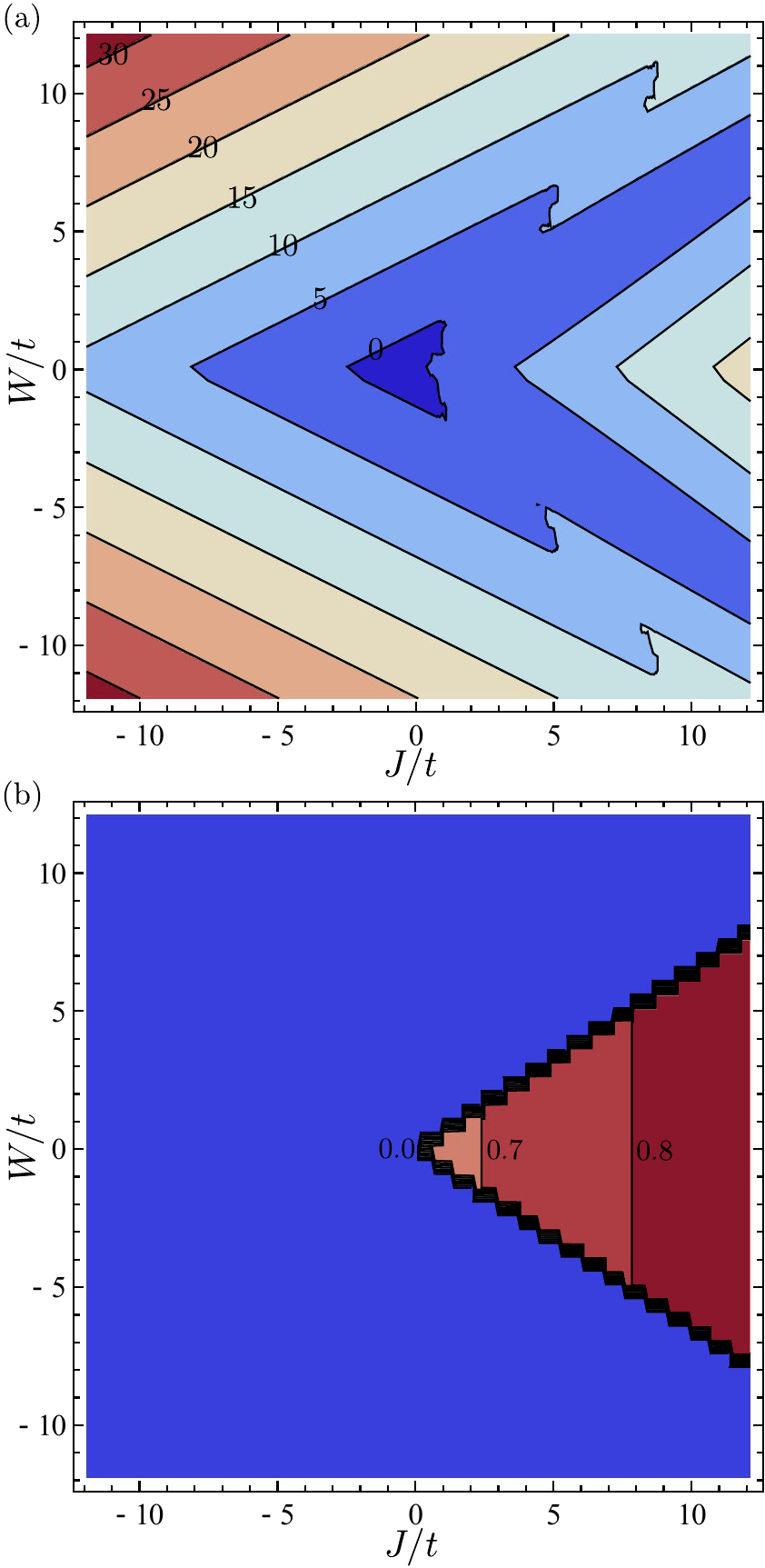} \ec
\caption{(Color online) Single-plaquette analysis for the the $t$-$J$-$W$ Hamiltonian with $J_\perp=J_z=J$. (a): Binding energy $\Delta_t$. Contours of constant $\Delta_t/t$ are labeled. (b): Matrix element $\abs{\Braket{4}{\dagg{\Delta}_d}{2}}$. Contours of constant $\abs{\Braket{4}{\dagg{\Delta}_d}{2}}$ are labeled.}
\label{t-Jperp-Jz-W_all}
\end{figure}

Next, we examine the Hamiltonian with independent $J_\perp=J_z=J$ and $W$ and zero $V$, the $t$-$J$-$W$ Hamiltonian. Contour plots for the binding energy $\Delta_t$ and $\abs{\Braket{4}{\dagg{\Delta}_d}{2}}$ for this Hamiltonian are shown in \fref{t-Jperp-Jz-W_all}. From these figures we see that $\Delta_t>0$ and $\abs{\Braket{4}{\dagg{\Delta}_d}{2}}\neq0$ for most values of $J/t>0$ and $W/t$ approximately between the lines $W/t=\pm 2J/3t$.

\begin{figure}
\bc \includegraphics[scale=.4]{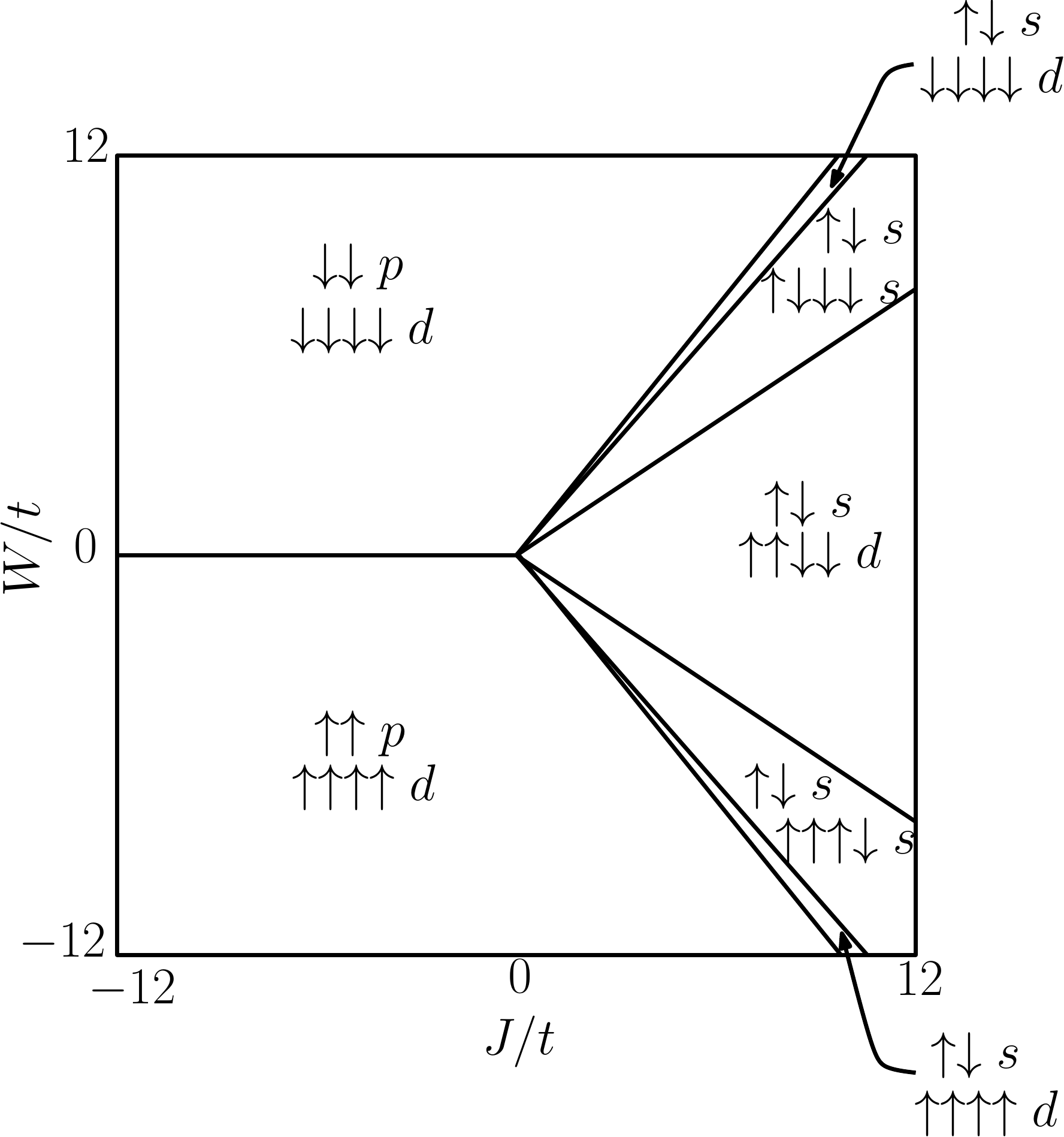} \ec
\caption{Configurations of the single-plaquette ground states for two and four molecules for the $t$-$J$-$W$ Hamiltonian with $J_\perp=J_z=J$.}
\label{configuration-t-Jperp-Jz-W}
\end{figure}

This behavior can be explained by considering the ground-state configurations and symmetries of $\ket{2}$ and $\ket{4}$ shown in \fref{configuration-t-Jperp-Jz-W}. If $W$ is large and positive, all of the spins will point down in the single-plaquette ground states: the $\ket{2}$ ground state is a $p$-wave $\ket{0,2}$ and the $\ket{4}$ ground state is a $d$-wave $\ket{0,4}$. For $W=0$ and $J/t>0$, it is energetically favorable to have an equal number of spin up and spin down molecules on each plaquette, and the $\ket{2}$ ground state is an $s$-wave $\ket{1,1}$, while the $\ket{4}$ ground state is a $d$-wave $\ket{2,2}$. As $W$ is decreased from a large positive value (and $J_\perp/t>0$), the $\ket{2}$ ground state first switches from $\ket{0,2}$ to an $s$-wave $\ket{1,1}$, and then the $\ket{4}$ ground state switches from $\ket{0,4}$ to an $s$-wave $\ket{1,3}$. As $W$ is further decreased, the $\ket{4}$ ground state switches to a $d$-wave symmetric $\ket{2,2}$. As $W$ is made large and negative, the process repeats with up spins replacing down spins. $\Braket{4}{\dagg{\Delta}_d}{2}$ vanishes for $W/t$ approximately outside the lines $W/t=\pm 2J/3t$ since the $\ket{4}$ ground state switches from $d$-wave to $s$-wave outside this region. In particular, $\Braket{4}{\dagg{\Delta}_d}{2}=0$ in the narrow region where the ground states are $s$-wave $\ket{1,1}$ and $d$-wave $\ket{1,3}$ because $\dagg{\Delta}_d$ creates one up spin and one down spin.

We do not consider Hamiltonians with nonzero $V$ here since they do not support $d$-wave superfluidity in our perturbative calculations as is described in the following section. We also note that, similar to the $t$-$J_\perp$ model where $\Delta_{40}$ becomes negative at large $J_\perp/t$, the region where a lattice of $\ket{2}$'s and $\ket{4}$'s is the true ground state is also limited in the models discussed in this section. However, as in the case of the $t$-$J_\perp$ model, the full phase diagrams discussed in Sec.\  \ref{sec:t-J-V-W_phase_diagrams} can still be accessed in experiments with ultracold polar molecules by adiabatic preparation starting with suitable initial states.

\subsection{First-Order Contributions to the $XXZ$ Effective Hamiltonian} \label{first-order-contributions}

In this section, we examine the question of which of the Hamiltonians defined by \eref{hamiltonian} have nonzero first-order contributions to the $XXZ$ effective Hamiltonian. A first-order shift in the ground-state energies, a nonzero $H_{1g}$ in \eref{effective_ham}, would contribute to, and generally dominate, the diagonal terms $f^{(4,2)}$, $f^{(2,2)}$, $f^{(4,4)}$, $h^{(4,2)}$, $h^{(2,2)}$, and $h^{(4,4)}$  
but not the off-diagonal term  $g^{(4,2)}$. Since the order $(t'/t)^2$ function $g^{(4,2)}$ is the only function contributing to $\tilde{J}_\perp$ [Eq.~(\ref{effective_strengths-a})], $\tilde{J}_\perp$ will generally be small compared to $\tilde{J}_{z1}$ and $\tilde{J}_{z2}$ [Eqs.~(\ref{effective_strengths-b}) and (\ref{effective_strengths-c})], if there are nonzero first-order contributions. Thus, this perturbative analysis predicts no robust superfluid phase if there are nonzero first-order contributions to the effective Hamiltonian. Therefore, we will not compute the phase diagram for those Hamiltonians that contribute first-order corrections. Note that there are no first-order shifts in the $t$-$J_\perp$ Hamiltonian studied in Sec.~\ref{sec:t-Jperp}.

To study when first-order contributions arise, we consider here the effects of the $J'_z$, $V'$, and $W'$ terms separately. Since these terms cannot couple $\ket{2}\ket{4}$ to $\ket{4}\ket{2}$, they will only contribute to the diagonal matrix elements of $H_{\t{eff}}$. Let $R$ and $R'$ label the plaquettes containing sites $r$ and $r'$, respectively, and let $n_{R\sigma}$ denote the number of molecules with spin $\sigma$ on plaquette $R$. Due to the $D_4$ symmetry, any single-plaquette eigenstate $\ket{n_{R\up},n_{R\dn}}$ that is non-degenerate within the manifold of states with constant $n_{R\up}$ and $n_{R\dn}$ satisfies \beq \Braket{n_{R\up},n_{R\dn}}{n_{r\sigma}}{n_{R\up},n_{R\dn}}= \frac{1}{4}n_{R\sigma}. \label{symm-condition} \eeq

First, consider the $V'$ term.
From \eref{symm-condition}, the state $\ket{n_R}\ket{n_{R'}}$ satisfies $$\ev{n_rn_{r'}}= \frac{1}{16}n_Rn_{R'},$$ where $n_r=n_{r\up}+n_{r\dn}$ and $n_R=n_{R\up}+n_{R\dn}$. Thus the $V'$ contributions vanish to first-order if and only if at least one of the two interacting plaquettes is empty. Since the effective Hamiltonian is constructed from $\ket{2}$ and $\ket{4}$ plaquettes, the $V'$ term will always contribute to first order and will not be considered further here. It is important to emphasize that our analysis should be considered to be exactly valid only in our perturbative regime, since it is believed that the $t$-$J$ model, which contains nonzero $V'$, supports $d$-wave superfluidity \cite{Lee2006,Ogata2008,Dagotto1994,Dagotto1993,Dagotto1992}.

Next, consider the $J'_z$ term.
Since $S_r^z=(n_{r\up}-n_{r\dn})/2$, the state $\ket{n_{R\up},n_{R\dn}}\ket{n_{R'\up},n_{R'\dn}}$ satisfies $$\ev{S_r^zS_{r'}^z} = \frac{1}{64}\lp n_{R\up}- n_{R\dn}\rp \lp n_{R'\up} - n_{R'\dn}\rp.$$ Thus, the $J'_z$ contributions vanish to first-order if and only if at least one of the two interacting plaquettes satisfies $n_\up=n_\dn$. In all of the regions identified in Sec.~\ref{sec:t-J-V-W:single_plaquette} as possibly supporting $d$-wave superfluidity in the $t$-$J_\perp$-$J_z$ Hamiltonian, $\ket{2}$ and $\ket{4}$ have this property. Therefore, $J'_z$ never contributes at first-order in the parameter regimes that we are interested in for this Hamiltonian.

Finally, consider the $W'$ term.
The state $\ket{n_{R\up},n_{R\dn}}\ket{n_{R'\up},n_{R'\dn}}$ satisfies $$\ev{n_rS_{r'}^z + n_{r'}S_r^z} = \frac{1}{16} (n_{R\up}n_{R'\up} - n_{R\dn}n_{R'\dn}).$$ Thus the $W'$ contributions vanish to first-order if and only if the two interacting plaquettes satisfy $n_{R\up}n_{R'\up} = n_{R\dn}n_{R'\dn}$. From \fref{configuration-t-Jperp-Jz-W}, we see that this condition is met for $\ket{2}\ket{4}$, $\ket{2}\ket{2}$ and $\ket{4}\ket{4}$ only for the region where the $\ket{2}$ ground state is $s$-wave symmetric $\ket{1,1}$ and the $\ket{4}$ ground state is $d$-wave symmetric $\ket{2,2}$. We also note that the condition $n_\up=n_\dn$, necessary for the $J'_z$ contributions to vanish at first-order, is satisfied by both $\ket{2}$ and $\ket{4}$ in this region. Notice, however, that the condition $n_\up = n_\dn$ is not satisfied by the $\ket{4}$ states outside this region. Thus, not only $W'$ but also $J'_z$ will give nonzero first-order contributions outside this region.

While other regions in \fref{configuration-t-Jperp-Jz-W} cannot exhibit a $d$-wave superfluid within our analysis, they may still exhibit interesting phases at appropriate filling fractions.
For example, the regions where the ground states are $\ket{1,1}$ and $\ket{0,4}$ or $\ket{4,0}$  might support a $d$-wave solid phase with an asymmetry between up and down spins, while  the regions where the ground states are $\ket{1,1}$ and $\ket{1,3}$ or $\ket{3,1}$ might support an $s$-wave solid phase. However, we will not discuss such phases further and will focus, instead, on the parameter space that has no first-order corrections and that is therefore capable of exhibiting $d$-wave superfluidity.

\subsection{Phase Diagrams} \label{sec:t-J-V-W_phase_diagrams}

\begin{figure}
\bc \includegraphics[scale=1]{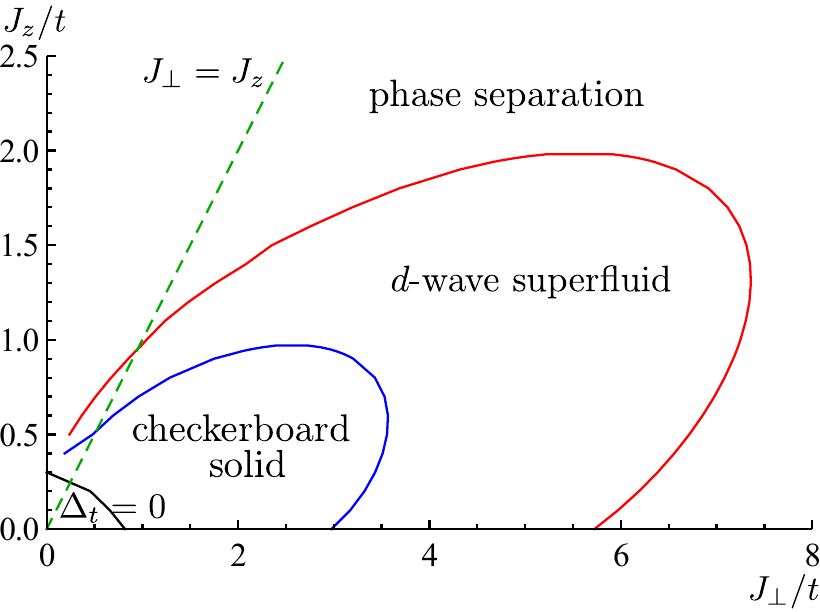} \ec
\caption{(Color online) Phase diagram for the $t$-$J_\perp$-$J_z$ Hamiltonian assuming $J'_\perp/t'=J_\perp/t$ and $J'_z/t'=J_z/t$. The region above the black curve is the region where $\Delta_t>0$. The SU(2)-symmetric Hamiltonian with $J_\perp=J_z$ is shown by the dashed green line. We do not compute the phase diagram along the $J_z/t$ axis since the $\ket{4}$ ground state is degenerate here.}
\label{phase-diagram-t-Jperp-Jz}
\end{figure}

We first consider the general $t$-$J_\perp$-$J_z$ Hamiltonian. There are four independent parameters, $J_\perp/t$, $J_z/t$, $J'_\perp/t'$, and $J'_z/t'$, so the phase diagram is four-dimensional. Notice that, within the perturbative treatment, the fifth parameter $t'/t$ affects only the overall energy scale but not the phase diagram. In order to plot a manageable phase diagram, we restrict the parameter space to $J'_\perp/t'=J_\perp/t$ and $J'_z/t'=J_z/t$. Using the expressions for $\tilde{J}_\perp$ and $\tilde{J}_{z1}$ given by \eref{effective_strengths} and the generalization of \eref{sw-functions-1} to include $J_z$ and $J'_z$, we find the phase boundaries by solving $\abs{\tilde{J}_\perp}=\abs{\tilde{J}_{z1}}$. The resulting phase diagram is shown in \fref{phase-diagram-t-Jperp-Jz}.  As is discussed in Sec.~\ref{sec:t-J-V-W:single_plaquette}, along the $J_z/t$ axis, the $\ket{4}$ ground state becomes doubly degenerate. Thus the effective Hamiltonian no longer maps to an $XXZ$ Hamiltonian, and we do not compute the phase diagram along this line. In the absence of the hardcore constraint, the $t$-$J_z$ Hamiltonian on a homogeneous lattice is studied in Ref.~\cite{Liu2011}. The $t$-$J_\perp$ phase diagram along the green dashed line $J'_\perp/t'=J_\perp/t$ shown in \fref{phase-diagram-t-Jperp} corresponds to the phase diagram along the $J_\perp/t$ axis in \fref{phase-diagram-t-Jperp-Jz}. For values of $J_z/t$ greater than about 1, the checkerboard solid phase is no longer present, and for values of $J_z/t$ greater than about 2, the superfluid phase is no longer present. We see that large values of $J_z$ have the effect of reducing the areas of the checkerboard solid and superfluid phases. However, by tuning $J_z/t$ to a value between about 1 and 2, the extent of the $d$-wave superfluid phase along the $J_\perp/t$ direction is increased relative to its value without the $J_z$ interaction. This is achieved by the suppression of the checkerboard solid and phase separation phases at small $J_\perp/t$ and large $J_\perp/t$, respectively.

\begin{figure}
\bc \includegraphics[scale=1]{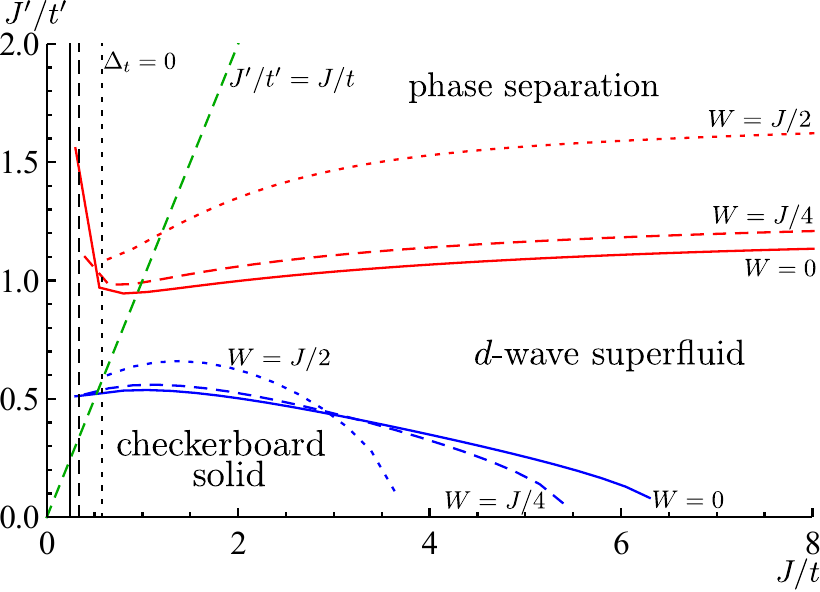} \ec
\caption{(Color online) Phase diagram for the $t$-$J$-$W$ Hamiltonian for $J_\perp=J_z=J$. Phase transitions for $W=0$ (the SU(2)-symmetric $t$-$J_\perp$-$J_z$ Hamiltonian) are shown in solid lines, transitions for $W=J/4$ are shown in dashed lines, and transitions for $W=J/2$ are shown in dotted lines. The regions to the right of the vertical black lines are the regions where $\Delta_t>0$ for these three values of $W$. The line $J'/t'=J/t$ is shown as a dashed green line.}
\label{phase-diagram-t-Jperp-Jz-W}
\end{figure}

We next consider the phase diagram for the SU(2)-symmetric $t$-$J_\perp$-$J_z$ Hamiltonian with $J_\perp=J_z=J$ shown in \fref{phase-diagram-t-Jperp-Jz-W} as solid lines. The phase diagram along the green dashed line $J'/t'=J/t$ in \fref{phase-diagram-t-Jperp-Jz-W} for the SU(2)-symmetric $t$-$J_\perp$-$J_z$ Hamiltonian corresponds to the phase diagram along the green dashed line $J_\perp=J_z$ in \fref{phase-diagram-t-Jperp-Jz}. The regions not in phase separation are reduced from those for the $t$-$J_\perp$ Hamiltonian (see \fref{phase-diagram-t-Jperp}). The fact that $J_z$ reduces the superfluid phase for the SU(2)-symmetric case is to be expected from \fref{phase-diagram-t-Jperp-Jz} since the line $J_\perp=J_z$ does not pass through the regions where the superfluid phase is enhanced by the presence of $J_z$.

Finally, we add $W$ and consider the phase diagram for the $t$-$J$-$W$ Hamiltonian with $J_\perp=J_z=J$ shown in \fref{phase-diagram-t-Jperp-Jz-W}. The phase diagram is shown along the lines $W=0$ (discussed above), $W=J/4$, and $W=J/2$, which are contained within the region where $\ket{2}$ is an $s$-wave symmetric $\ket{1,1}$ and $\ket{4}$ is a $d$-wave symmetric $\ket{2,2}$ (see \fref{configuration-t-Jperp-Jz-W}). Increasing $W$ moves the transition between the superfluid and phase separation phases up along the $J'/t'$ axis and therefore slightly increases the region of $d$-wave superfluidity. Increasing $W$ also decreases the extent of the checkerboard solid phase along the $J/t$ axis.

In summary, within our treatment, Hamiltonians involving $V$ do not support superfluidity. At the same time, we identify regions of parameter space where $J_z$ and $W$ enhance the $d$-wave superfluid phase.

\section{Experimental Realizations of the Perturbative Calculation} \label{sec:experimental_realizations}

The perturbative results obtained above 
can be regarded only as a qualitative guess as to the behavior of the simplest homogeneous square lattice since intraplaquette and interplaquette couplings are equal in this case. Therefore, in this section, we propose experimental configurations accurately described by our perturbative analysis. In Sec.~\ref{sec:stack}, we describe a one-dimensional stack of plaquettes. Then, in Sec.~\ref{sec:checkerboard_splitting}, we briefly describe an experimentally more challenging two-dimensional configuration.

\subsection{One-Dimensional Stack of Plaquettes} \label{sec:stack}

\begin{figure}
\bc \includegraphics[scale=.7]{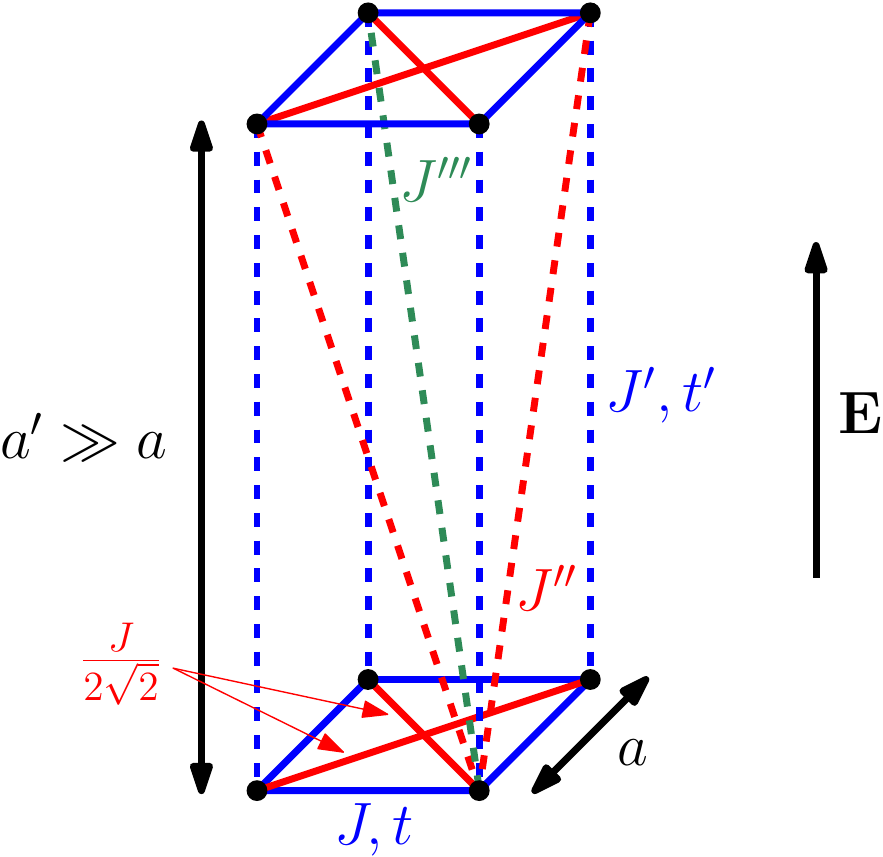} \ec
\caption{(Color online) Geometry of the one-dimensional stack of plaquettes. The external dc electric field is perpendicular to the plaquettes.}
\label{stacked-plaquettes-geometry}
\end{figure}

As shown in Fig.\ \ref{stacked-plaquettes-geometry}, we propose to apply a dc electric field in the vertical ($\hat{\mathbf{z}}$) direction. We then propose to stack plaquettes with sides of length $a$ on top of each other along $\hat{\mathbf{z}}$ a distance $a'$ apart. The strength of dipolar interactions between two molecules is proportional to \beq \frac{1-3\cos^2\theta}{r^3}, \label{dipolar-strength} \eeq where $r$ is the distance between the molecules and $\theta$ is the angle made between the line connecting the molecules and the external dc electric field. Due to the $1/r^3$ dependence, interplaquette dipolar interactions can be treated perturbatively relative to the intraplaquette interactions provided $a' \gg a$.
One way of achieving this geometry experimentally is to use different frequency lasers to create the optical lattices in the plane and in the vertical direction. The distances $a$ and $a'$ can also be controlled by varying the angle at which the lasers interfere \cite{Nelson2007,Peil2003} or by holographic techniques \cite{Curtis2002}.

The plaquettes can be made by interfering lasers of wavelength $2a$ and $4a$ to create a superlattice \cite{Trotzky2008} such that tunneling between plaquettes in a plane is negligible. To avoid in-plane interplaquette dipole-dipole interactions, molecules in neighboring stacks may have to be removed. The required addressability can be acheived by applying temporary additional light shifts or electric field gradients. Alternatively, instead of emptying neighboring stacks, an extreme version of the superlattice can be used to separate the stacks enough to make both the tunneling and the dipolar interactions between them negligible.

As is shown in \fref{stacked-plaquettes-geometry}, with this geometry, there is a hopping amplitude $t$ between nearest neighbors within a plaquette and a hopping amplitude $t'$ along $\hat{\mathbf{z}}$ between nearest-neighbor plaquettes. The amplitudes $t$ and $t'$ can be controlled separately by varying the intensities of the lasers making the lattices in each direction. There are five strengths, two within a plaquette and three between plaquettes, of the dipolar interactions $J_\perp$, $J_z$, $V$, and $W$ discussed here generically as $J$. As shown in \fref{stacked-plaquettes-geometry}, let $J$ be the strength of dipolar interactions between nearest neighbors within a plaquette, and let $J'$, $J''$, and $J'''$ be the strengths of the dipolar interactions between plaquettes. The ratios $J'/J$, $J''/J$, and $J'''/J$ are functions of the ratio $a'/a$ and are controlled separately from the ratio $t'/t$. For $a'>a$, by \eref{dipolar-strength}, $J'$, $J''$, and $J'''$  are the opposite sign of $J$.

\subsubsection{Phase Diagrams}

For a one-dimensional stack of plaquettes, we calculate the same phase diagrams computed above for the two-dimensional lattice of plaquettes. In an experiment with one independent dipolar interaction strength $J$, there are three independent parameters: $J/t$, $a'/a$, and $t'/t$ or, equivalently, $J/t$, $J'/t'$, and $a'/a$. In order to compare with the phase diagrams computed above, we choose the latter set of parameters. It is currently possible to use lasers of wavelength $1064\t{ nm}$ to produce optical lattices with spacing $a'= 532\t{ nm}$. Assuming that it is also possible to use a second wavelength in the range $400-600 \t{ nm}$, in the following phase diagrams, we use $a'/a=5/2$. For $a'/a=5/2$, $J'/J=-0.128$, $J''/J=-0.081$, and $J'''/J=-0.054$.

\begin{figure}
\bc \includegraphics[scale=1]{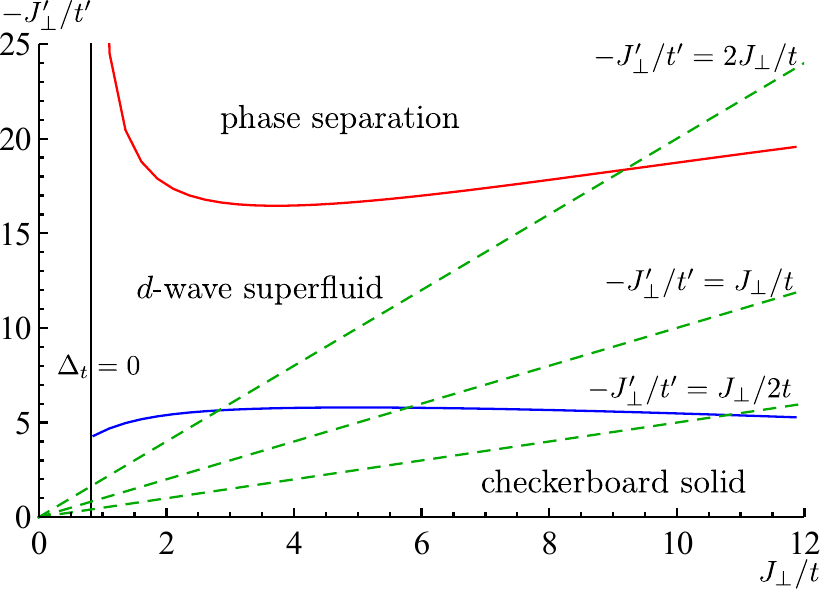} \ec
\caption{(Color online) Phase diagram for the $t$-$J_\perp$ Hamiltonian for a one-dimensional stack of plaquettes with $a'/a=5/2$. Regions to the right of the vertical black line are where $\Delta_t>0$. The lines $-J'_\perp/t'=J_\perp/2t$, $-J'_\perp/t'=J_\perp/t$, and $-J'_\perp/t'=2J_\perp/t$ are shown by the green dashed lines.}
\label{phase-diagram-t-Jperp-stacked}
\end{figure}

The phase diagram for the $t$-$J_\perp$ Hamiltonian is shown in \fref{phase-diagram-t-Jperp-stacked}. Since it is possible with this scheme to control $J'_\perp/t'$ by varying $a'/a$ and $t'/t$, we plot the lines $-J'_\perp/t'=J_\perp/2t$, $-J'_\perp/t'=J_\perp/t$, and $-J'_\perp/t'=2J_\perp/t$ in \fref{phase-diagram-t-Jperp-stacked}. This phase diagram is qualitatively similar to the corresponding phase diagram for the two-dimensional case (\fref{phase-diagram-t-Jperp}). However, the overall vertical scale of the phase diagram is increased for the stack resulting in a larger region of $d$-wave superfluidity. The increase in the vertical scale can be explained by the presence of the $J''_\perp$ and $J'''_\perp$ terms, which reduce the $f_\perp$ functions relative to the $f_t$ and $g_t$ functions in \eref{sw-functions-1}. Thus, a larger value of $J'_\perp/t'$ is needed to reach the phase boundaries. We refer the reader to Appendix~\ref{sec:scale_increase} for a symmetry-based explanation for why $J''_\perp$ and $J'''_\perp$ reduce the $f_\perp$ functions. Similar arguments hold for the increased scales in the $t$-$J_\perp$-$J_z$ and the $t$-$J$-$W$ phase diagrams discussed below.

\begin{figure}
\bc \includegraphics[scale=1]{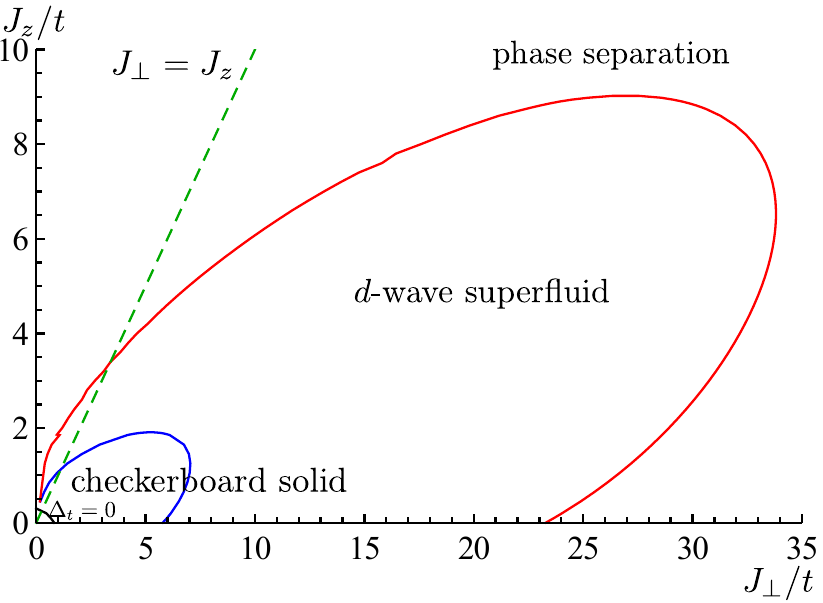} \ec
\caption{(Color online) Phase diagram for the $t$-$J_\perp$-$J_z$ Hamiltonian assuming $J'_\perp/t'=J_\perp/t$ and $J'_z/t'=J_z/t$ for a one-dimensional stack of plaquettes with $a'/a=5/2$. The region above the black curve is the region where $\Delta_t>0$. The SU(2)-symmetric Hamiltonian with $J_\perp=J_z$ is shown by the dashed green line.}
\label{phase-diagram-t-Jperp-Jz-stacked}
\end{figure}

The phase diagram for the $t$-$J_\perp$-$J_z$ Hamiltonian with $J'_\perp/t'=J_\perp/t$ and $J'_z/t'=J_z/t$ is shown in \fref{phase-diagram-t-Jperp-Jz-stacked}. This phase diagram is qualitatively similar to the corresponding phase diagram for the two-dimensional case shown in \fref{phase-diagram-t-Jperp-Jz}. As with the two-dimensional case, $J_z$ can increase the superfluid phase.

\begin{figure}
\bc \includegraphics[scale=1]{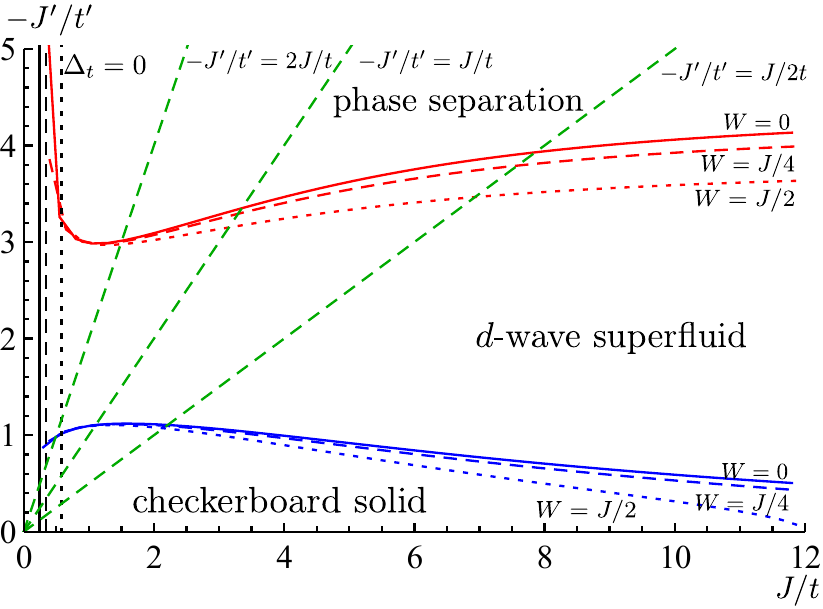} \ec
\caption{(Color online) Phase diagram for the $t$-$J$-$W$ Hamiltonian for $J_\perp=J_z=J$ for a one-dimensional stack of plaquettes with $a'/a=5/2$. Phase transitions for $W$=0 (the SU(2)-symmetric $t$-$J_\perp$-$J_z$ Hamiltonian) are shown in solid lines, transitions for $W=J/4$ are shown in dashed lines, and transitions for $W=J/2$ are shown in dotted lines. The regions to the right of the vertical black lines are the regions where $\Delta_t>0$ for these three values of $W$. The lines $-J'/t'=J/2t$, $-J'/t'=J/t$, and $-J'/t'=2J/t$ are shown by the green dashed lines.}
\label{phase-diagram-t-J-W-stacked}
\end{figure}

The phase diagram for the SU(2)-symmetric $t$-$J_\perp$-$J_z$ Hamiltonian is shown in \fref{phase-diagram-t-J-W-stacked} as solid lines. Again, the diagram is qualitatively similar to its counterpart for the two-dimensional system (\fref{phase-diagram-t-Jperp-Jz-W}) with the range of $J'/t'$ not in the phase separation regime increased.

The overall scale of the phase diagram in \fref{phase-diagram-t-Jperp-Jz-stacked} is increased from the corresponding diagram for the two-dimensional system shown in \fref{phase-diagram-t-Jperp-Jz} as is expected from Figs.~\ref{phase-diagram-t-Jperp-stacked} and \ref{phase-diagram-t-J-W-stacked}. The $t$-$J_\perp$ phase diagram along the middle green dashed line $-J'_\perp/t'=J_\perp/t$ shown in \fref{phase-diagram-t-Jperp-stacked} corresponds to the phase diagram along the $J_\perp/t$ axis in \fref{phase-diagram-t-Jperp-Jz-stacked}. The SU(2)-symmetric phase diagram along the middle green dashed line $-J'/t'=J/t$ shown in \fref{phase-diagram-t-J-W-stacked} corresponds to the phase diagram along the green dashed line $J_\perp=J_z$ in \fref{phase-diagram-t-Jperp-Jz-stacked}. Since both of these green lines in Figs.~\ref{phase-diagram-t-Jperp-stacked} and \ref{phase-diagram-t-J-W-stacked} pass through larger regions of the superfluid phase than in the corresponding phase diagrams for the two dimensional plane (Figs.~\ref{phase-diagram-t-Jperp} and \ref{phase-diagram-t-Jperp-Jz-W}), the overall scale of the $t$-$J_\perp$-$J_z$ phase diagram (\fref{phase-diagram-t-Jperp-Jz-stacked}) increases. The superfluid phase can be further increased by decreasing the slope of the line relating $J'_\perp/t'$ ($J'/t'$) and $J_\perp/t$ ($J/t$) in \fref{phase-diagram-t-Jperp-stacked} (\fref{phase-diagram-t-J-W-stacked}) since lines with shallower slopes pass through larger regions of the superfluid phase.

The phase diagram for the $t$-$J$-$W$ Hamiltonian for $J_\perp=J_z=J$ is shown in \fref{phase-diagram-t-J-W-stacked}. In contrast to the phase diagram for the two-dimensional $t$-$J$-$W$ Hamiltonian, in the stack geometry, an increase in $W$ moves the transition between the superfluid and phase separation phases down rather than up, thus reducing the superfluid phase. Furthermore, in the stack geometry, an increase in $W$ only slightly decreases the extent of the checkerboard solid phase. Thus for the stack of plaquettes, the $W$ term can slightly increase the superfluid region only for small $\abs{J'/t'}$.

\subsubsection{Preparation and Detection}
\label{sec:preparation_detection}

Since there are no strong relaxation mechanisms in optical lattice experiments using cold atoms and molecules, it is not an easy task to prepare the ground state of a given Hamiltonian. One strategy is to prepare an easier state that is the ground state of another Hamiltonian and to adiabatically change that Hamiltonian to the desired one \cite{Sorensen2010,Trebst2006,Duan2003,Trotzky2008}. Here we consider the adiabatic preparation of the $d$-wave superfluid phase at the point where $\tilde{J}_{z1}=0$ for the $t$-$J_\perp$ Hamiltonian. This corresponds to preparing the ground state of the $XX$ magnet. We propose first adiabatically preparing $\ket{2}$ and $\ket{4}$ states alternating in a stack and then adiabatically preparing the ground state of the $t$-$J_\perp$ Hamiltonian. The preparation of the ground state of the $XX$ magnet is similar to the method described in \cite{Sorensen2010} for adiabatically preparing the ground state of the Heisenberg antiferromagnet.

First, consider bringing the plaquettes far apart so as to avoid interplaquette interactions and preparing the $\ket{2}$ and $\ket{4}$ ground states alternating along the stack. Let $H_0$ be the $t$-$J_\perp$ Hamiltonian on a single plaquette and consider the following single-plaquette Hamiltonians 
\beqnn
H_2(\tau)\e H_0 - B(\tau)\lp S_1^z - S_3^z\rp, \label{fc2}\\
H_4(\tau)\e H_0 - B(\tau)\lp S_1^z + S_2^z - S_3^z - S_4^z\rp, \label{fc4}
\eeqnn 
used to prepare the $\ket{2}$ and $\ket{4}$ ground states, respectively. Here $B(\tau)$ is the strength of an effective alternating magnetic field on the sites of the plaquette as a function of time $\tau$. For large positive values of $B(\tau)$, the ground state of $H_2$ for two molecules is up on 1 and down on 3, while the ground state of $H_4$ for four molecules is up on 1 and 2 and down on 3 and 4. These states can be prepared using single-site addressability provided by electric field gradients or high-resolution optics \cite{Simon2011,Sherson2010}. For values of $J_\perp/t$ of interest, in a plaquette with two molecules, the energy gap between the ground state and first excited state of $H_2$ never closes as $B(\tau)$ is reduced to zero. The same is true of $H_4$ for a plaquette with four molecules. Thus the $\ket{2}$ and $\ket{4}$ ground states of the $t$-$J_\perp$ Hamiltonian can be prepared by adiabatically reducing the effective magnetic field $B(\tau)$ from some large initial value to zero in $H_2$ and $H_4$, respectively. Once the $\ket{2}$ and $\ket{4}$ ground states have been prepared, the plaquettes can be brought to a distance $a'$ from each other by changing the angle at which the lasers interfere or by holographic techniques.

Once the $\ket{2}$ and $\ket{4}$ ground states have been prepared, the ground state of the $t$-$J_\perp$ Hamiltonian at the point $\tilde{J}_{z1}=0$ can be prepared by adiabatically turning off the effective magnetic field $\tilde{B}(\tau)$ acting on the effective spins $\ket{\Up}$ and $\ket{\Dn}$ in the Hamiltonian \begin{multline} H(\tau) = \tilde{J}_\perp \sum_{R}\lp S^X_R S^X_{R+1} + S^Y_RS^Y_{R+1}\rp\\
+ \tilde{B}(\tau) \sum_{R} (-1)^R S_R^Z. \label{prep-hamiltonian} \end{multline} The first term is just the $XXZ$ effective Hamiltonian \eref{xxz-hamiltonian} at the point $\tilde{J}_{z1}=0$, while the second term describes an alternating effective magnetic field between plaquettes along the $Z$ direction. The energy gap between the ground state and the first excited state of the Hamiltonian \eref{prep-hamiltonian} decreases monotonically with decreasing magnetic field but never closes, so the $d$-wave superfluid phase can, in principle, be prepared by adiabatically turning off $\tilde{B}(\tau)$. However, for a stack of $N$ plaquettes, the energy gap between the ground state and the first excited state of the $XX$ chain is proportional to $1/N$. Although it is outside the scope of the present paper, the ideal $\tilde{B}(\tau)$ over a given time $T$ can be calculated by mapping \eref{prep-hamiltonian} to free fermions and maximizing the superfluid order parameter when $\tilde{B}(T)=0$. Since the energy gap is large for large $\tilde{B}(\tau)$ and only scales as $1/N$ for small $\tilde{B}(\tau)$, it is likely that this ideal $\tilde{B}(\tau)$ decreases rapidly initially and more slowly at later times.

As a rough estimate for the minimum time $T$ necessary to prepare the superfluid phase at the point $\tilde{J}_{z1}=0$, we estimate $T=N/\tilde{J}_\perp$. The strength of the dipole-dipole interactions at a distance of $200\t{ nm}$ is roughly $J_\perp\approx 2\pi\cross 4 \t{ kHz}$ for the molecule KRb. The value of $J'_\perp\approx -2\pi\cross 500\t{ Hz}$ is then fixed by the ratio $a'/a$. $t$ is chosen such that $\tilde{J}_\perp$ is as large as possible while perturbation theory is valid. $t'$ is chosen such that $\tilde{J}_{z1}=0$. Then for $J_\perp/t=4$, $t\approx 2\pi\cross 1 \t{ kHz}$ and $\tilde{J}_{z1}=0$ requires $t'\approx 2\pi\cross 300 \t{ Hz}.$ At this point, $\tilde{J}_\perp\approx 2\pi\cross 1.5\t{ Hz}$ so $T\approx 100N\t{ ms}.$ For these parameters, $\Delta_t\approx 2\pi\cross 2\t{ kHz}$ and $\Delta_\perp\approx 2\pi\cross 4\t{ kHz}$ so the conditions $t'\ll \Delta_t$ and $J'_\perp\ll \Delta_\perp$ are met, and perturbation theory is valid. Using the inverse gap at $B(\tau)=0$ in Eqs.~(\ref{fc2}) and (\ref{fc4}) gives $400\t{ }\mu\t{s}$ as a rough estimate for the minimum time necessary to prepare the $\ket{2}$ and $\ket{4}$ ground states. Thus, as expected, the preparation time is dominated by the preparation of the ground state of the $XX$ chain. Therefore, assuming an optimistic coherence time of $1\t{ s}$, roughly $10$ plaquettes can be prepared. For LiCs, $J_\perp\approx 2\pi\cross 400\t{ kHz}$ at a distance of $200\t{ nm}$ allowing roughly $10^3$ plaquettes to be prepared in $1\t{ s}$. These numbers can be improved by reducing the ratio $a'/a$ or by preparing a superfluid away from the point $\tilde{J}_{z1}=0$.

The effective magnetic fields in Eqs.~(\ref{fc2})--(\ref{prep-hamiltonian}) can be created using tensor shifts \cite{Wall2010,Brennen2007,Wall2009,Micheli2007,Kotochigova2010,Gorshkov2011} and superlattices \cite{Trotzky2008} so that up and down spins (both the initial $\ket{\up}$ and $\ket{\dn}$ and the effective $\ket{\Up}$ and $\ket{\Dn}$) have different potential energies.

The $d$-wave superfluid phase can be detected via second-order noise correlations in the expanding molecular cloud which is proportional to the the four-point function  $$G_{\sigma\sigma'}(\v{Q},\v{Q'})\propto \ev{n_\sigma(\v{Q})n_{\sigma'}(\v{Q'})} - \ev{n_\sigma(\v{Q})} \ev{n_{\sigma'}(\v{Q'})}$$ at the time the molecules are released from the trap \cite{Altman2004,Spielman2007,Rey2009}. Here $$n_\sigma(\v{Q})\propto \sum_{r,r'} \ee^{\ii \v{Q}\cdot\v{L}_{rr'}} \dagg{c}_{r\sigma} c_{r'\sigma}$$ is the quasi-momentum distribution and $\v{L}_{rr'}$ is the vector connecting lattice sites $r$ and $r'$. Since $G_{\up\dn}(\v{Q},\v{Q'})$ contains terms proportional to $\ev{\dagg{\Delta}_{d,R} \Delta_{d,R'}}$ (where $R$ and $R'$ label plaquettes), a $d$-wave superfluid will exhibit interference fringes at $\v{Q}+\v{Q'}= 2\pi m\rf{z}/a'$ for any integer (half-integer) $m$ where $\hat{\mathbf{z}}$ is the unit vector along the stack when $\tilde{J}_\perp<0$ ($>0$).
Since the state has $d_{x^2-y^2}$ symmetry, these fringes will be modulated in the $x$-$y$ plane by an envelope that vanishes along the nodal lines $Q_x=\pm Q_y$ and $Q'_x=\pm Q'_y$ \cite{Rey2009}. Therefore, the behavior of the fringes will be similar to that shown in Ref.~\cite{Rey2009}, except that the superfluid phase will be signaled by fringes along $\rf{z}$ instead of fringes in the $x$-$y$ plane. Due to the absence of cycling transitions in molecules, the noise correlation measurements may have to be done by first converting the molecules back into atoms \cite{Ni2008}.

\subsection{Two-Dimensional Realization of the Perturbative Regime} \label{sec:checkerboard_splitting}

While we have discussed the one-dimensional stack of plaquettes, $d$-wave superfluidity is associated with a plane. An experiment observing $d$-wave superfluidity in a two-dimensional lattice would therefore be more relevant. The simple solution of increasing the spacing between plaquettes in a two-dimensional lattice to reduce the dipolar interactions is not feasible since this would typically reduce the tunneling amplitude $t'$ between plaquettes to essentially zero.

However, if the energy difference between the rotor levels is not constant throughout a lattice of plaquettes, then the $J_\perp$ dipolar interaction between plaquettes could be suppressed \footnote{We thank M.\ Lewenstein for pointing this out to us.}. This can be achieved by introducing an extra splitting $\Delta$ between rotor levels in some plaquettes and not in others. Then the Hamiltonian for each plaquette with the extra splitting gets an extra term $$\Delta\sum_r S_r^z,$$ where the sum is taken over the four sites of the plaquette. Let $(R_x,R_y)$ be the integer coordinates of the plaquette labeled by $R$. Suppose that the splitting is arranged in a checkerboard fashion, so that for integers $n$ and $m$ plaquettes $(R_x+2n,R_y+2m)$ get the extra splitting $\Delta$ while plaquettes $(R_x+2n+1,R_y+2m+1)$ do not. In this case, dipolar interactions between nearest-neighbor plaquettes will be suppressed by $\Delta$ since they are off resonant; however, dipolar interactions between next-nearest-neighbor (diagonal) plaquettes will not be suppressed since they will stay resonant. Thus to make the perturbative calculations valid, four separate splittings would need to be introduced to suppress both nearest and next-nearest-neighbor interactions. However, with only one splitting $\Delta$, a one-dimensional chain of plaquettes can be simulated experimentally.

Finally, for $\Delta$ that is large enough to make perturbation theory valid, a lattice of plaquettes filled with $|1,1\rangle$ and $|2,2\rangle$ states will typically be an excited-state of the system of decoupled plaquettes, and the $d$-wave superfluid state will typically be an excited state of the full Hamiltonian of weakly coupled plaquettes. Nevertheless, such states can still be prepared adiabatically from appropriate excited states.

\section{Conclusions} \label{sec:conclusions}

We have shown that the $t$-$J_\perp$ Hamiltonian on a square lattice, in the regime of weakly coupled plaquettes, exhibits a $d$-wave superfluid phase in addition to the checkerboard solid phase and phase separation. The addition of large $J_z$ or $W$ interactions destroys the superfluid phase; however, we have identified ranges of these parameters, for which the superfluid phase is enhanced. Any nonzero $V$ destroys the superfluid phase in this perturbative analysis.

These perturbative calculations can be used as a qualitative guess for the behavior of the simplest experiments, which are outside of the perturbative limit. Furthermore, the perturbative regime can be accessed in experiments in a one-dimensional stack of plaquettes. The phase diagrams for the one-dimensional stack of plaquettes are qualitatively similar to the phase diagrams for the two-dimensional lattice of plaquettes. By experimentally observing the phase diagrams in both the perturbative and non-perturbative regimes for the one-dimensional stack, one may be able to understand the relationship between the calculations presented here and the non-perturbative phase diagrams. This knowledge may then be useful in understanding the relationship between the two-dimensional phase diagrams presented here and experiments on a homogeneous two-dimensional lattice. Similar results might be achievable experimentally with the large-magnetic-moment atoms dysprosium \cite{Lu2011} and chromium \cite{Griesmaier2005} instead of polar molecules.

Ultracold polar molecules have the potential for experimentally observing $d$-wave superfluidity in a controlled environment that could allow us to learn about the physics of the $t$-$J$ model. We hope that the insight gained by these investigations could help to explain the physics of high-temperature superconductivity and result in many theoretical and practical applications.

\begin{acknowledgements}

We thank John Preskill, Maciej Lewenstein, Rajdeep Sensarma, Salvatore Manmana, Kaden Hazzard, Mikhail Lukin, Eugene Demler, Netanel Lindner, Norbert Schuch, Steven Flammia, Spyridon Michalakis, Gang Chen, Michael Foss-Feig, and Xiao Yin for discussions. This work was supported by the Rose Hills Foundation, the Lee A. DuBridge Foundation, the NSF (PFC and Grants No.~PHY-0803371 and PIF-0904017), and ARO with funding from the DARPA OLE program.

\end{acknowledgements}

\appendix

\section{Group Theoretic Techniques} \label{sec:group_theory}

In this appendix, we describe the group theoretic techniques that we have used to diagonalize the Hamiltonians and to study their symmetries. Exact diagonalization \cite{Tinkham1964} is discussed in general in Appendix~\ref{sec:group_theory:diagonalization}, while the symmetries \cite{Isaev2010,Yao2007} of the irreducible representations of $D_4$ are discussed in Appendix~\ref{representations-description}. Finally, in Appendix~\ref{sec:scale_increase}, we use symmetry arguments to explain why the overall scale of the phase diagrams in the stack geometry is larger than the corresponding scale in the two-dimensional geometry. Related group theoretic techniques are discussed in the context of exact diagonalization of the Hubbard Hamiltonian in Refs.~\cite{Schumann2002,Fano1992}.

\subsection{Diagonalization of the Hamiltonian} \label{sec:group_theory:diagonalization}

As is discussed in Sec.~\ref{sec:t-Jperp:single_plaquette}, the operators $n_{\up}$ and $n_{\dn}$ commute with the Hamiltonian \eref{hamiltonian} and with each other, so we diagonalize subspaces with fixed values of $n_{\up}$ and $n_{\dn}$ separately. The Hamiltonian \eref{hamiltonian} has the symmetries of a square described by the group $D_4$ so we use the irreducible representations of $D_4$ to further simplify the task of diagonalizing each subspace. The discussion below uses $D_4$ as an example but is general and can be applied to any finite group. In this section, we closely follow Ref.~\cite{Tinkham1964}.

$D_4$ has $h=8$ group elements corresponding to the symmetries of a square, considered here to be lying in the $xy$ plane: the identity $E$, rotations by $\pi$ around the $x$, $y$, and $z$ axes $C_{2x}$, $C_{2y}$, and $C_{2z}$, rotations by $\pi$ around the lines $x=y$ and $x=-y$ $C_{2xy}$ and $C_{2x\bar{y}}$, and counterclockwise and clockwise rotations by $\pi/2$ around the $z$ axis $C_{4z}$ and $C_{4z}^{-1}$. There are five conjugacy classes and thus five irreducible representations for $D_4$. The five classes are the identity $E$, the $\pi$ rotation about the $z$ axis $C_2$ consisting of $C_{2z}$, the two $\pi/2$ rotations $2C_4$ consisting of $C_{4z}$ and $C_{4z}^{-1}$, the two $\pi$ rotations about the $x$ and $y$ axes $2C'_2$ consisting of $C_{2x}$ and $C_{2y}$, and the two $\pi$ rotations about the lines $x=y$ and $x=-y$ $2C''_2$ consisting of $C_{2xy}$ and $C_{2x\bar{y}}$. There are four one-dimensional representations $A_1$, $A_2$, $B_1$, and $B_2$ and one two-dimensional representation $E$. The character table for $D_4$ is shown in \tref{character-table}.

\begin{table}
\bc
\begin{tabular}{|r|rrrrr||l|}
\hline
 & $E$ & $C_2$ & $2C_4$ & $2C'_2$ & $2C''_2$ & Symmetries\\
\hline
$A_1$ & 1 & 1 & 1 & 1 & 1 & $s$\\
$A_2$ & 1 & 1 & 1 & $-1$ & $-1$ & $s$\\
$B_1$ & 1 & 1 & $-1$ & 1 & $-1$ & $d_{x^2-y^2}$\\
$B_2$ & 1 & 1 & $-1$ & $-1$ & 1 & $d_{xy}$\\
$E$ & 2 & $-2$ & 0 & 0 & 0 & $p_x$ and $p_y$\\
\hline
\end{tabular}
\ec
\caption{Character table for the group $D_4$ taken from Ref.~\cite{Tinkham1964}, along with the classification of each representation as is discussed in Sec.~\ref{representations-description}.} \label{character-table}
\end{table}

The number of times the $n$th irreducible representation appears in the decomposition of a reducible representation is \cite{Tinkham1964} \beq a_n=\frac{1}{h}\sum_R \star{\chi^{(n)}(R)}\chi(R), \label{representation} \eeq where the sum is taken over all group elements $R$, $\chi^{(n)}(R)$ is the character of $R$ in the $n$th irreducible representation, and $\chi(R)$ is the character of $R$ in the reducible representation. Let $\ket{\alpha}$ denote a basis function in the occupation basis. By writing the eight symmetry operations in the occupation basis for a fixed $n_\up$ and $n_\dn$ subspace of the full Hilbert space, we calculate the character of each element in this occupation representation. Using \eref{representation}, we calculate the group structure of each of these subspaces. The results are shown in \tref{group-structure}.

\begin{table}
\bc
\begin{tabular}{|r|r|r|}
\hline
$n_\up,n_\dn$ & Dimension & Representation\\
\hline
1,0 & 4 & $A_1\oplus B_2\oplus E$\\
2,0 & 6 & $A_2\oplus B_2\oplus 2E$\\
1,1 & 12 & $2A_1\oplus A_2\oplus B_1\oplus 2B_2\oplus 3E$\\
3,0 & 4 & $A_2\oplus B_1\oplus E$\\
2,1 & 12 & $A_1\oplus 2A_2\oplus 2B_1\oplus B_2\oplus 3E$\\
4,0 & 1 & $B_1$\\
3,1 & 4 & $A_2\oplus B_1\oplus E$\\
2,2 & 6 & $A_1\oplus A_2\oplus 2B_1 \oplus E$\\
\hline
\end{tabular}
\ec
\caption{Group structure of the single-plaquette Hilbert space with $D_4$ symmetry in the presence of $n_\up$ and $n_\dn$ conservation. The results are symmetric on interchange of $n_\up$ and $n_\dn$.}
\label{group-structure}
\end{table}

As we will see below, the Hamiltonian is block diagonal in the representation basis, so we would like to change basis from the occupation basis $\ket{\alpha}$ to the representation basis $\ket{\phi_{i\lambda}^{(n)}}$ to reduce the dimensions of the matrices needing to be diagonalized. Here $\ket{\phi_{i\lambda}^{(n)}}$ refers to a basis function for the $i$th row of the $n$th irreducible representation. If the $n$th irreducible representation is present more than once in the decomposition of the occupation representation, then there will be $a_n>1$ orthogonal basis functions for the same row of the $n$th irreducible representation indexed by $\lambda$. Let there be $c$ irreducible representations and $c$ classes. (For $D_4$, $c=5$). $\ket{\alpha}$ can be expanded in terms of the $\ket{\phi_{i\lambda}^{(n)}}$ as \beq \ket{\alpha}=\sum_{n=1}^c\sum_{i=1}^{l_n}\sum_{\lambda=1}^{a_n} b_{i\lambda}^{(n)} \ket{\phi_{i\lambda}^{(n)}}, \label{occupation-decomposition} \eeq where $l_n$ is the dimension of the $n$th irreducible representation.

The $\ket{\phi_{i\lambda}^{(n)}}$ and $b_{i\lambda}^{(n)}$ are found by applying the projection operator \beq P_{ij}^{(n)} = \frac{l_n}{h}\sum_R \star{\Gamma^{(n)}(R)}_{ij} P_R \label{projector}\eeq to $\ket{\alpha}$. $\Gamma^{(n)}(R)$ is the $n$th irreducible representation of the group element $R$ and $P_R$ is the operator corresponding to $R$ that acts on functions instead of coordinates and satisfies $$P_Rg(x)=g(R^{-1}x).$$ To compute $P_R$ in practice, one finds the action of $R$ on the spatial indicies of the creation operators used to define the second-quantized wave function.

The projector $P_{ii}^{(n)}$ projects into the $i$th row of the $n$th irreducible representation, so \beq\ket{f_{i\lambda(\alpha)}^{(n)}}=P_{ii}^{(n)}\ket{\alpha} \label{representation-projection} \eeq yields a function $\ket{f_{i\lambda(\alpha)}^{(n)}}$ that transforms as the $i$th row of the $n$th irreducible representation. For the $i$th row of the $n$th irreducible representation, a given $\ket{\alpha}$ will only be composed of a function belonging to one of the $a_n$ copies, the $\lambda(\alpha)$th copy, of the $i$th row of the $n$th irreducible representation even if $a_n>1$. Note that $\ket{f_{i\lambda(\alpha)}^{(n)}}$ will be zero if $a_n=0$. For each nonzero $\ket{f_{i\lambda(\alpha)}^{(n)}}$, the corresponding normalized basis function is, up to a global phase, $$\ket{\phi_{i\lambda(\alpha)}^{(n)}}= \lp\braket{f_{i\lambda(\alpha)}^{(n)}}{f_{i\lambda(\alpha)}^{(n)}}\rp^{-1/2} \ket{f_{i\lambda(\alpha)}^{(n)}}.$$ The coefficients $b_{i\lambda}^{(n)}$ of the decomposition \eref{occupation-decomposition} are $$b_{i\lambda}^{(n)}=\begin{cases} 0 \t{ if } a_n=0\\
\lp\braket{f_{i\lambda(\alpha)}^{(n)}}{f_{i\lambda(\alpha)}^{(n)}}\rp^{1/2} \delta_{\lambda,\lambda(\alpha)} \t{ if } a_n\neq 0 \end{cases}.$$

To find all of the basis functions for the representation basis, we compute \eref{representation-projection} for every basis function $\ket{\alpha}$ in the occupation basis for all $P_{ii}^{(n)}$ \footnote{If $l_n>1$, once the basis function for a single row, the $i$th row, of the $n$th irreducible representation has been found by \eref{representation-projection}, it is also possible to find the basis functions for the other $l_n-1$ rows by applying the projection operator to the basis function just found by the property $\big|\phi_{j\lambda}^{(n)}\textrm{$\big\rangle$} = P_{ji}^{(n)} \big|\phi_{i\lambda}^{(n)}\textrm{$\big\rangle$}$.}. By applying the same projection operator to different functions $\ket{\alpha}$, different functions for the same row of the same representation will be generated for all $\lambda=1,\dots,a_n$. Once all of the $\ket{\phi_{i\lambda}^{(n)}}$ have been found, the Hamiltonian can be transformed into the representation basis by $$H_{\t{representation}}= \dagg{S}H_{\t{occupation}}S,$$ where $S$ is the transformation matrix given by $S_{\alpha,\phi}=\braket{\alpha}{\phi_{i\lambda}^{(n)}}$. Since $P_R$ commutes with $H$ for all $R$, \beqn \Braket{\phi_{i\lambda(\alpha)}^{(n)}}{H}{\phi_{j\mu(\beta)}^{(m)}} &\propto& \Braket{\alpha}{P_{ii}^{(n)}HP_{jj}^{(m)}}{\beta}\\
\e \Braket{\alpha}{HP_{ii}^{(n)}P_{jj}^{(m)}}{\beta} \propto \delta_{n,m}\delta_{i,j}. \eeqn Thus, matrix elements between different irreducible representations or different rows within the same irreducible representation vanish. Therefore, this transformation into the representation basis can be made separately for each row of each irreducible representation, and each of the resulting Hamiltonians can be diagonalized separately. This greatly reduces the dimension of the Hamiltonians that need to be diagonalized since the dimension of the Hamiltonian for the $i$th row of the $n$th irreducible representation is $a_n$. States found in the representation basis can be transformed back into the occupation basis by $$\braket{\alpha}{\psi}=\sum_{\phi} S_{\alpha,\phi} \braket{\phi}{\psi}.$$

In order to compute the projector \eref{projector}, it is necessary to know the representation $\Gamma^{(n)}(R)$ of each group element $R$ in each representation $n$. For the one-dimensional representations, the representations $\Gamma^{(n)}(R)$ are just the characters $\chi^{(n)}(R)$ listed in \tref{character-table}. In the basis $\bbmat x\\
y\ebmat$, the representations of the group elements in the $E$ representation are \begin{multline} \Gamma^{(E)}(E)=\bbmat 1 & 0\\
0 & 1\ebmat \qquad \Gamma^{(E)}(C_{2z})=\bbmat -1 & 0\\
0 & -1\ebmat\\
\Gamma^{(E)}(C_{4z})=\bbmat 0 & -1\\
1 & 0\ebmat \qquad \Gamma^{(E)}(C_{4z}^{-1})=\bbmat 0 & 1\\
-1 & 0\ebmat\\
\Gamma^{(E)}(C_{2x})= \bbmat 1 & 0\\
0 & -1 \ebmat \qquad \Gamma^{(E)}(C_{2y})=\bbmat -1 & 0\\
0 & 1\ebmat\\
\Gamma^{(E)}(C_{2xy})=\bbmat 0 & 1\\
1 & 0\ebmat \qquad \Gamma^{(E)}(C_{2x\bar{y}})=\bbmat 0 & -1\\
-1 & 0\ebmat. \label{Erepresentations}
\end{multline}

\subsection{Classification of the Eigenstates} \label{representations-description}

In this section, we classify the symmetries of the five representations of $D_4$ and give examples of states transforming as each of these representations. Since the $A$ and $B$ representations are one-dimensional, we can study their symmetries directly from the character table \tref{character-table}. The $A$ states are symmetric under $\pi/2$ rotations $2C_4$, while the $B$ states are antisymmetric under these rotations. For this reason, the $A$ states are classified as $s$-wave and the $B$ states are classified as $d$-wave. The $A_1$ representation is symmetric under all five classes, while the $A_2$ representation is antisymmetric under $2C'_2$ and $2C''_2$. The $B_1$ representation is symmetric under $2C'_2$ and is antisymmetric under $2C''_2$, while the $B_2$ representation is antisymmetric under $2C'_2$ and is symmetric under $2C''_2$. Thus, $B_1$ is classified as $d_{x^2-y^2}$ and $B_2$ is classified as $d_{xy}$.

The $E$ representation is two-dimensional, so we need to consider the representations of the elements given by \eref{Erepresentations}. Both rows of the representation are antisymmetric under $\pi$ rotations about the $z$ axis. For this reason, $E$ is classified as $p$-wave. The first row transforms into the second row and the second row transforms into negative the first row under a positive rotation by $\pi/2$ about the $z$ axis. The first row is antisymmetric and the second row is symmetric under a $\pi$ rotation about the $x$ axis. The first row transforms into the second row and the second row transforms into the first row under a $\pi$ rotation about the line $y=x$. Similar considerations for the other rotations lead to the classification of the first row of the $E$ representation as $p_x$ and the second row as $p_y$.

Simple examples of states in the $A_1$, $B_2$, and $E$ representations come from $n_\up=1$ and $n_\dn=0$. An example of an $A_1$ state is $$\ket{A_1}=\frac{1}{2}\lp \ket{\up000}+\ket{0\up00}+ \ket{00\up0} +\ket{000\up}\rp.$$ We use the notation in which, for example,
$$\ket{\up0\dn0}=\dagg{c}_{1\up}\dagg{c}_{3\dn}\ket{0},$$ and the numbering of sites within a plaquette is given in \fref{plaquette-geometry}. This state is clearly invariant under all five of the $D_4$ classes. An example of a $B_2$ state is $$\ket{B_2}=\frac{1}{2}\lp \ket{\up000}-\ket{0\up00}+\ket{00\up0} -\ket{000\up}\rp.$$ This state is invariant under $C_2$ and $2C''_2$ but changes sign under $2C_4$ and $2C'_4$. An example of two $E$ states is \beqn \ket{E_x}\e \frac{1}{2}\lp\ket{\up000} -\ket{0\up00} -\ket{00\up0} + \ket{000\up}\rp,\\
\ket{E_y}\e \frac{1}{2}\lp\ket{\up000} + \ket{0\up00} -\ket{00\up0} -\ket{000\up}\rp.\eeqn These states are antisymmetric under $\pi$ rotations about $z$. Under a positive rotation by $\pi/2$ about $z$, $\ket{E_x}\to\ket{E_y}$ and $\ket{E_y}\to-\ket{E_x}$. Under a $\pi$ rotation about the $x$ axis, $\ket{E_x}\to\ket{E_y}$ and $\ket{E_y}\to-\ket{E_x}$. Under a $\pi$ rotation about the line $y=x$, $\ket{E_x}\to\ket{E_y}$ and $\ket{E_y}\to\ket{E_x}$. Thus $\ket{E_x}$ transforms as $p_x$ and $\ket{E_y}$ transforms as $p_y$.

A simple example of an $A_2$ state comes from $n_\up=2$ and $n_\dn=0$ where \beqn \ket{A_2}\e\frac{1}{2}\lp \ket{\up\up00}+ \ket{0\up\up0}+ \ket{00\up\up} -\ket{\up00\up}\rp\\
\e \frac{1}{2}\lp \dagg{c}_{1\up}\dagg{c}_{2\up} + \dagg{c}_{2\up}\dagg{c}_{3\up} +\dagg{c}_{3\up}\dagg{c}_{4\up} + \dagg{c}_{4\up}\dagg{c}_{1\up}\rp\ket{0}.\eeqn This state is invariant under $C_2$ and $2C_4$ since this corresponds to cyclically permuting the indices. Under $2C'_2$ and $2C''_2$, $\ket{A_2}\to-\ket{A_2}$ since this corresponds to swapping indices.

A simple example of a $B_1$ state is just the state from $n_\up=4$ and $n_\dn=0$ $$\ket{B_1}=\ket{\up\up\up\up}=\dagg{c}_{1\up} \dagg{c}_{2\up} \dagg{c}_{3\up} \dagg{c}_{4\up}\ket{0}.$$ As can be seen by appropriately switching the indicies on the creation operators, this state has the $B_1$ symmetries discussed above.

\subsection{Scale Increase in the Phase Diagrams for the Stack Geometry}
\label{sec:scale_increase}

In this section, we use symmetry arguments to explain the increase in the overall scale of phase diagrams for the stack geometry (Figs.~\ref{phase-diagram-t-Jperp-stacked}, \ref{phase-diagram-t-Jperp-Jz-stacked}, and \ref{phase-diagram-t-J-W-stacked}) relative to the corresponding diagrams for the two-dimensional geometry (Figs.~\ref{phase-diagram-t-Jperp}, \ref{phase-diagram-t-Jperp-Jz}, and \ref{phase-diagram-t-Jperp-Jz-W}). The main reason for this increase is that the $J'_\perp$, $J'_z$, and $W'$ transition matrix elements between the low- and high-energy subspaces decrease as $J''/J'$ and $J'''/J'$ are varied from $0$ to $1$. As a result, all phase transitions occur at smaller values of $t'$. To gain intuition for why these matrix elements decrease as $J''$ and $J'''$ approach $J'$, we study the $t$-$J_\perp$ Hamiltonian for the stack geometry in the regime where $J'_\perp=J''_\perp=J'''_\perp$ (obtained as $a'/a\to\infty$). A similar argument holds for the other Hamiltonians considered.

Consider two neighboring plaquettes in a stack in the $\ket{2,2}\ket{2,2}$ state coupled by the $J'_\perp$ perturbing Hamiltonian $H_{1\perp}$ with $J'_\perp=J''_\perp=J'''_\perp$. From \tref{groundstates}, both plaquettes have $B_1$ symmetry. Since every vertex in one of the plaquettes is equally coupled to every vertex in the other plaquette, $H_{1\perp}$ is symmetric under arbitrary permutations of the vertices within one of the plaquettes. As a result, the matrix element $$\bra{1,3}\bra{3,1}H_{1\perp} \ket{2,2}\ket{2,2}$$ will vanish unless both $\ket{1,3}$ and $\ket{3,1}$ have $B_1$ symmetry; otherwise, an appropriate permutation can be used to show that this matrix element is equal to negative itself. This drastically reduces the magnitude of $f_\perp^{(4,4)}$ relative to its value in the $J''_\perp=J'''_\perp=0$ case. A similar argument shows that $f_\perp^{(4,2)}$ and $f_\perp^{(2,2)}$ vanish for $J'_\perp=J''_\perp=J'''_\perp$ (since the $\ket{1,3}$ and $\ket{3,1}$ manifolds have no $A_1$ states. See \tref{group-structure}).

\section{Basis Vectors in the Representation Basis} \label{sec:rep_basis}

In this appendix, we list the basis vectors for the irreducible representations for $D_4$. The results are symmetric on interchange of up spins with down spins. We use the notation $\ket{n_\up,n_\dn(\Gamma)}$ to refer to the basis vector for a plaquette with $n_\up$ up spins and $n_\dn$ down spins in the irreducible representation $\Gamma$. If there are $a_n>1$ copies of an irreducible representation, we denote them by a superscript. We use subscripts $x$ and $y$ to denote the first and second rows of the $E$ representation, respectively. Thus $\ket{1,1(E_y^1)}$ refers to the basis vector for $\ket{1,1}$ in the first copy of the second row of the $E$ representation and $\ket{2,1(B_1^2)}$ refers to the basis vector for $\ket{2,1}$ in the second copy of the $B_1$ representation.

\subsubsection{$n_\up=1$ and $n_\dn=0$}

\noindent $\ds\ket{1,0(A_1)}= \frac{1}{2}\lp \ket{\up000} + \ket{0\up00} + \ket{00\up0} + \ket{000\up} \rp$\\
$\ds\ket{1,0(B_2)}= \frac{1}{2}\lp \ket{\up000} - \ket{0\up00} + \ket{00\up0} - \ket{000\up} \rp$\\
$\ds\ket{1,0(E_x)}= \frac{1}{2}\lp \ket{\up000} - \ket{0\up00} - \ket{00\up0} + \ket{000\up} \rp$\\
$\ds\ket{1,0(E_y)}= \frac{1}{2}\lp \ket{\up000} + \ket{0\up00} - \ket{00\up0} - \ket{000\up} \rp$

\subsubsection{$n_\up=2$ and $n_\dn=0$}

\noindent $\ds\ket{2,0(A_2)}= \frac{1}{2}\lp \ket{\up\up00} + \ket{0\up\up0} - \ket{\up00\up} + \ket{00\up\up} \rp$\\
$\ds\ket{2,0(B_2)}= \frac{1}{2}\lp \ket{\up\up00} - \ket{0\up\up0} + \ket{\up00\up} + \ket{00\up\up} \rp$\\
$\ds\ket{2,0(E_x^1)}= \frac{1}{\sqrt{2}}\lp \ket{\up\up00} - \ket{00\up\up} \rp$\\
$\ds\ket{2,0(E_y^1)}= \frac{1}{\sqrt{2}}\lp \ket{0\up\up0} + \ket{\up00\up} \rp$\\
$\ds\ket{2,0(E_x^2)}= \frac{1}{\sqrt{2}}\lp \ket{\up0\up0} - \ket{0\up0\up} \rp$\\
$\ds\ket{2,0(E_y^2)}= \frac{1}{\sqrt{2}}\lp \ket{\up0\up0} + \ket{0\up0\up} \rp$

\subsubsection{$n_\up=1$ and $n_\dn=1$}

\noindent $\ds\ket{1,1(A_1^1)}= \frac{1}{2\sqrt{2}}( \ket{\dn\up00} - \ket{\up\dn00} + \ket{0\dn\up0} - \ket{0\up\dn0}\\
\hspace{1cm} + \ket{\dn00\up} + \ket{00\dn\up} - \ket{\up00\dn} - \ket{00\up\dn})$\\
$\ds\ket{1,1(A_1^2)}= \frac{1}{2}\lp \ket{\dn0\up0} - \ket{\up0\dn0} + \ket{0\dn0\up} - \ket{0\up0\dn}\rp$\\
$\ds\ket{1,1(A_2)}= \frac{1}{2\sqrt{2}}( \ket{\dn\up00} + \ket{\up\dn00} + \ket{0\dn\up0} + \ket{0\up\dn0} \\
- \ket{\dn00\up} + \ket{00\dn\up} - \ket{\up00\dn} + \ket{00\up\dn})$\\
$\ds\ket{1,1(B_1)}= \frac{1}{2\sqrt{2}}( \ket{\dn\up00} - \ket{\up\dn00} - \ket{0\dn\up0} + \ket{0\up\dn0}\\
 - \ket{\dn00\up} + \ket{00\dn\up} + \ket{\up00\dn} - \ket{00\up\dn})$\\
$\ds\ket{1,1(B_2^1)}= \frac{1}{2\sqrt{2}}( \ket{\dn\up00} + \ket{\up\dn00} - \ket{0\dn\up0} - \ket{0\up\dn0}\\
 + \ket{\dn00\up} + \ket{00\dn\up} + \ket{\up00\dn} + \ket{00\up\dn})$\\
$\ds\ket{1,1(B_2^2)}= \frac{1}{2}\lp \ket{\dn0\up0} - \ket{\up0\dn0} - \ket{0\dn0\up} + \ket{0\up0\dn}\rp$\\
$\ds\ket{1,1(E_x^1)}= \frac{1}{2}\lp \ket{\dn\up00} + \ket{\up\dn00} - \ket{00\dn\up} - \ket{00\up\dn}\rp$\\
$\ds\ket{1,1(E_y^1)}= \frac{1}{2}\lp \ket{0\up\dn0} +\ket{0\dn\up0} + \ket{\up00\dn} + \ket{\dn00\up}\rp$\\
$\ds\ket{1,1(E_x^2)}= \frac{1}{2}\lp \ket{0\up\dn0} - \ket{0\dn\up0} - \ket{\up00\dn} + \ket{\dn00\up}\rp$\\
$\ds\ket{1,1(E_y^2)}= \frac{1}{2}\lp \ket{\dn\up00} - \ket{\up\dn00} - \ket{00\dn\up} + \ket{00\up\dn}\rp$\\
$\ds\ket{1,1(E_x^3)}= \frac{1}{2}\lp \ket{\dn0\up0} + \ket{\up0\dn0} - \ket{0\dn0\up} - \ket{0\up0\dn}\rp$\\
$\ds\ket{1,1(E_y^3)}= \frac{1}{2}\lp \ket{\dn0\up0} + \ket{\up0\dn0} + \ket{0\dn0\up} + \ket{0\up0\dn}\rp$

\subsubsection{$n_\up=3$ and $n_\dn=0$}

\noindent $\ds\ket{3,0(A_2)}= \frac{1}{2}\lp \ket{\up\up\up0} + \ket{\up\up0\up} + \ket{\up0\up\up} + \ket{0\up\up\up}\rp$\\
$\ds\ket{3,0(B_1)}= \frac{1}{2}\lp \ket{\up\up\up0} - \ket{\up\up0\up} + \ket{\up0\up\up} - \ket{0\up\up\up}\rp$\\
$\ds\ket{3,0(E_x)}= \frac{1}{2}\lp \ket{\up\up\up0} + \ket{\up\up0\up} - \ket{\up0\up\up} - \ket{0\up\up\up}\rp$\\
$\ds\ket{3,0(E_y)}= \frac{1}{2}\lp \ket{\up\up\up0} - \ket{\up\up0\up} - \ket{\up0\up\up} + \ket{0\up\up\up}\rp$

\subsubsection{$n_\up=2$ and $n_\dn=1$}

\noindent $\ds\ket{2,1(A_1)} = \frac{1}{2\sqrt{2}}( \ket{\dn\up\up0} - \ket{\up\up\dn0} - \ket{\up\dn0\up} - \ket{\dn0\up\up}\\
 + \ket{0\dn\up\up} + \ket{\up0\dn\up} + \ket{\up\up0\dn} - \ket{0\up\up\dn})$\\
$\ds\ket{2,1(A_2^1)} = \frac{1}{2\sqrt{2}}( \ket{\dn\up\up0} + \ket{\up\up\dn0} + \ket{\up\dn0\up} + \ket{\dn0\up\up}\\
 + \ket{0\dn\up\up} + \ket{\up0\dn\up} + \ket{\up\up0\dn} + \ket{0\up\up\dn})$\\
$\ds\ket{2,1(A_2^2)} = \frac{1}{2}( \ket{\up\dn\up0} + \ket{\dn\up0\up} + \ket{0\up\dn\up} + \ket{\up0\up\dn})$\\
$\ds\ket{2,1(B_1^1)} = \frac{1}{2\sqrt{2}}( \ket{\dn\up\up0} + \ket{\up\up\dn0} - \ket{\up\dn0\up} + \ket{\dn0\up\up}\\
 - \ket{0\dn\up\up} + \ket{\up0\dn\up} - \ket{\up\up0\dn} - \ket{0\up\up\dn})$\\
$\ds\ket{2,1(B_1^2)} = \frac{1}{2}( \ket{\up\dn\up0} - \ket{\dn\up0\up} - \ket{0\up\dn\up} + \ket{\up0\up\dn})$\\
$\ds\ket{2,1(B_2)} = \frac{1}{2\sqrt{2}}( \ket{\dn\up\up0} - \ket{\up\up\dn0} + \ket{\up\dn0\up} - \ket{\dn0\up\up}\\
 - \ket{0\dn\up\up} + \ket{\up0\dn\up} - \ket{\up\up0\dn} + \ket{0\up\up\dn})$\\
$\ds\ket{2,1(E_x^1)} = \frac{1}{2}( \ket{\dn\up\up0} + \ket{\up\dn0\up} - \ket{\up0\dn\up} - \ket{0\up\up\dn})$\\
$\ds\ket{2,1(E_y^1)} = \frac{1}{2}( \ket{\up\up\dn0} - \ket{\dn0\up\up} + \ket{0\dn\up\up} - \ket{\up\up0\dn})$\\
$\ds\ket{2,1(E_x^2)} = \frac{1}{2}( \ket{\up\up\dn0} - \ket{\dn0\up\up} - \ket{0\dn\up\up} + \ket{\up\up0\dn})$\\
$\ds\ket{2,1(E_y^2)} = \frac{1}{2}( \ket{\dn\up\up0} - \ket{\up\dn0\up} - \ket{\up0\dn\up} + \ket{0\up\up\dn})$
$\ds\ket{2,1(E_x^3)} = \frac{1}{2}( \ket{\up\dn\up0} + \ket{\dn\up0\up} - \ket{0\up\dn\up} - \ket{\up0\up\dn})$\\
$\ds\ket{2,1(E_y^3)} = \frac{1}{2}( \ket{\up\dn\up0} - \ket{\dn\up0\up} + \ket{0\up\dn\up} - \ket{\up0\up\dn})$\\

\subsubsection{$n_\up=4$ and $n_\dn=0$}

\noindent$\ds\ket{4,0(B_1)}= \ket{\up\up\up\up}$

\subsubsection{$n_\up=3$ and $n_\dn=1$}

\noindent $\ds\ket{3,1(A_2)}= \frac{1}{2}\lp \ket{\dn\up\up\up} - \ket{\up\dn\up\up} + \ket{\up\up\dn\up} - \ket{\up\up\up\dn} \rp$\\
$\ds\ket{3,1(B_1)}= \frac{1}{2}\lp \ket{\dn\up\up\up} + \ket{\up\dn\up\up} + \ket{\up\up\dn\up} + \ket{\up\up\up\dn} \rp$\\
$\ds\ket{3,1(E_x)}= \frac{1}{2}\lp \ket{\dn\up\up\up} - \ket{\up\dn\up\up} - \ket{\up\up\dn\up} + \ket{\up\up\up\dn} \rp$\\
$\ds\ket{3,1(E_y)}= -\frac{1}{2}\lp \ket{\dn\up\up\up} + \ket{\up\dn\up\up} - \ket{\up\up\dn\up} - \ket{\up\up\up\dn} \rp$

\subsubsection{$n_\up=2$ and $n_\dn=2$}

\noindent $\ds\ket{2,2(A_1)}= \frac{1}{2}\lp \ket{\dn\dn\up\up} - \ket{\up\dn\dn\up} - \ket{\dn\up\up\dn} + \ket{\up\up\dn\dn} \rp$\\
$\ds\ket{2,2(A_2)}= \frac{1}{\sqrt{2}}\lp \ket{\dn\up\dn\up} - \ket{\up\dn\up\dn} \rp$\\
$\ds\ket{2,2(B_1^1)}= \frac{1}{2}\lp \ket{\dn\dn\up\up} + \ket{\up\dn\dn\up} + \ket{\dn\up\up\dn} + \ket{\up\up\dn\dn} \rp$\\
$\ds\ket{2,2(B_1^2)}= \frac{1}{\sqrt{2}}\lp \ket{\dn\up\dn\up} + \ket{\up\dn\up\dn} \rp$\\
$\ds\ket{2,2(E_x)}= \frac{1}{\sqrt{2}} \lp \ket{\up\dn\dn\up} - \ket{\dn\up\up\dn}\rp$\\
$\ds\ket{2,2(E_y)}= \frac{1}{\sqrt{2}} \lp \ket{\dn\dn\up\up} - \ket{\up\up\dn\dn}\rp$

\section{Explicit Expressions for Important Ground States and Energies} \label{sec:ground_states}

In this appendix, we give explicit expressions for the ground states and energies for those $\ket{2}$ and $\ket{4}$ states responsible for $d$-wave superfluidity in the $t$-$J_\perp$ Hamiltonian, $t$-$J_\perp$-$J_z$ Hamiltonian, and the $t$-$J$-$W$ Hamiltonian with $J_\perp=J_z=J$. We also discuss intuition for the symmetries of some of these states.

\subsection{$t$-$J_\perp$ Hamiltonian}\label{ground-states-tJperp}

Here we give the $\ket{2}$ and $\ket{4}$ ground states of the $t$-$J_\perp$ Hamiltonian responsible for $d$-wave superfluidity (see \tref{groundstates} and \fref{t-Jperp_all}). The $s$-wave symmetric $\ket{1,1}$ ground state for $J_\perp/t > -1.22$ is 
\begin{equation}
\ket{1,1(A_1)}  \propto b\ket{1,1(A_1^1)} + \ket{1,1(A_1^2)} \label{eqa1}
\end{equation}
 with energy \begin{multline} E_g(1,1)=-\frac{1}{16} \Bigg[\sqrt{\left(18-8 \sqrt{2}\right) \lp \jpot\rp^2+2048}\\
+\left(4+\sqrt{2}\right) \jpot \Bigg], \end{multline} 
 where
$$b= \frac{1}{32}\lb\sqrt{(9-4\sqrt{2})\lp\jpot\rp^2 +1024}+ (2\sqrt{2}-1)\jpot\rb.$$

The $d$-wave symmetric $\ket{2,2}$ ground state for $J_\perp/t>0$ is
\begin{equation}
\ket{2,2(B_1)} \propto \frac{\sqrt{65}-1}{8}\ket{2,2(B_1^1)} - \ket{2,2(B_1^2)}
\end{equation} with energy $$E_g(2,2)= \frac{1-\sqrt{65}}{4\sqrt{2}} \frac{J_\perp}{t}.$$

We can understand the symmetries of the $\ket{2}$ ground states as follows. In the limit of vanishing $t$, the ground state will be an eigenstate of the $J_\perp$ interaction. The two eigenstates of the $J_\perp$ interaction for an up and a down molecule on sites $r$ and $r'$ are \beq\ket{\pm}_{rr'} = \frac{1}{\sqrt{2}} \lp \dagg{c}_{r\up} \dagg{c}_{r'\dn} \pm \dagg{c}_{r\dn} \dagg{c}_{r'\up}\rp \ket{0} \label{pm-def} \eeq since 
$$\frac{J_\perp}{2}\lp S_r^+S_{r'}^- + S_r^-S_{r'}^+\rp \ket{\pm}_{rr'}= \pm\frac{J_\perp}{2}\ket{\pm}_{rr'}.$$ (Note that $\ket{-}_{rr'}= \dagg{s}_{rr'}\ket{0}$ where $\dagg{s}_{rr'}$ is given by \eref{singlet-def}). Thus for positive $J_\perp$, the singlet state $\ket{-}$ is the ground state of the $J_\perp$ interaction. Note that it is disadvantageous to have identical spins since, in that case, the $J_\perp$ interaction vanishes. For two molecules with small tunneling amplitude $t$, the $\ket{2}$ ground state is thus a superposition of singlets on the four nearest-neighbor bonds. Diagonalizing the effective Hamiltonian of this system shows that the ground state is the following symmetric superposition: $$\ket{-}_{12} + \ket{-}_{23} + \ket{-}_{34} + \ket{-}_{41}.$$ This state has $s$-wave symmetry and is proportional to the basis vector $\ket{1,1(A_1)}$, consistent with Eq.\ (\ref{eqa1}). For negative $J_\perp$, the $\ket{+}$ state is the ground state of the $J_\perp$ interaction on two sites, and the two resulting single-plaquette ground states have $p$-wave symmetry.

The symmetries of the $\ket{4}$ ground states can be understood by considering only nearest-neighbor interactions. In this case, for positive $J_\perp$, the ground state is $$\ket{2,2(B_1^2)}-\ket{2,2(B_1^1)},$$ which has $d_{x^2-y^2}$ symmetry. For negative $J_\perp$, the ground state is $$\ket{2,2(B_1^2)}+\ket{2,2(B_1^1)},$$ which also has $d_{x^2-y^2}$ symmetry. When the effects of next-nearest neighbors are considered, the coefficients in the superposition of $\ket{2,2(B_1^1)}$ and $\ket{2,2(B_1^2)}$ are changed.

\subsection{$t$-$J_\perp$-$J_z$ Hamiltonian}

Here we give the $\ket{2}$ and $\ket{4}$ ground states of the $t$-$J_\perp$-$J_z$ Hamiltonian responsible for $d$-wave superfluidity (see \fref{t-Jperp-Jz_all}). The $s$-wave symmetric $\ket{1,1}$ ground state is \begin{equation} \ket{1,1(A_1)} \propto a\ket{1,1(A_1^1)} + \ket{1,1(A_1^2)}\end{equation} with energy \begin{multline} E_g(1,1)= -\frac{1}{16} \Bigg[\sqrt{2048-2 \left(4 \sqrt{2}-9\right) \lp\frac{J_\perp}{t}+2 \frac{J_z}{t}\rp^2}\\
+\left(4+\sqrt{2}\right) \frac{J_\perp}{t}+2 \left(4+\sqrt{2}\right) \frac{J_z}{t}\Bigg], \non \end{multline} where \begin{multline} a = \frac{1}{32\sqrt{2}}\Bigg[\sqrt{2048-2 \left(4 \sqrt{2}-9\right) \lp\frac{J_\perp}{t}+2 \frac{J_z}{t}\rp^2}\\
-\left(\sqrt{2}-4\right) \frac{J_\perp}{t}-2 \left(\sqrt{2}-4\right) \frac{J_z}{t} \Bigg]. \non \end{multline}

The $d$-wave symmetric $\ket{2,2}$ ground state is \begin{equation} \ket{2,2(B_1)} \propto b\ket{2,2(B_1^1)} + \ket{2,2(B_1^2)} \end{equation} with energy \begin{multline} E_g(2,2)= \frac{1}{8} \Bigg\{-\Bigg[130 \lp\frac{J_\perp}{t}\rp^2+16 \left(2 \sqrt{2}-1\right) \frac{J_\perp}{t} \frac{J_z}{t}+\\
32 \left(9-4 \sqrt{2}\right) \lp\frac{J_z}{t}\rp^2\Bigg]^{1/2} +\sqrt{2} \frac{J_\perp}{t}-16 \frac{J_z}{t}\Bigg\}, \non \end{multline} where \begin{multline} b = \frac{1}{8} \frac{t}{J_\perp} \Bigg\{ -\Bigg[65 \lp\frac{J_\perp}{t}\rp^2+8 \left(2 \sqrt{2}-1\right) \frac{J_\perp}{t} \frac{J_z}{t}\\
+16 \left(9-4 \sqrt{2}\right) \lp\frac{J_z}{t}\rp^2\Bigg]^{1/2}+\frac{J_\perp}{t}+\left(8 \sqrt{2}-4\right) \frac{J_z}{t} \Bigg\}. \non \end{multline}

\subsection{$t$-$J$-$W$ Hamiltonian with $J_\perp=J_z=J$}

Here we give the $\ket{2}$ and $\ket{4}$ ground states of the $t$-$J$-$W$ Hamiltonian with $J_\perp=J_z=J$ responsible for $d$-wave superfluidity (see Figs.~\ref{t-Jperp-Jz-W_all} and \ref{configuration-t-Jperp-Jz-W}). The $s$-wave symmetric $\ket{1,1}$ ground state is \begin{equation} \ket{1,1(A_1)} \propto a\ket{1,1(A_1^1)} + \ket{1,1(A_1^2)} \end{equation} with energy
\begin{multline} E_g(1,1)= -\frac{1}{16} \Bigg[ \sqrt{18 \left(9-4 \sqrt{2}\right) \lp\frac{J}{t}\rp^2+2048}+\\
3 \left(4+\sqrt{2}\right) \frac{J}{t} \Bigg], \end{multline}  where $$a= \frac{1}{32} \lb\sqrt{9 \left(9-4 \sqrt{2}\right) \lp\frac{J}{t}\rp^2+1024}+\left(6 \sqrt{2}-3\right) \frac{J}{t}\rb.$$

The $d$-wave symmetric $\ket{2,2}$ ground state is \begin{equation} \ket{2,2(B_1)} \propto b\ket{2,2(B_1^1)} + \ket{2,2(B_1^2)}\end{equation} with energy
$$E_g(2,2)= \frac{1}{8} \frac{J}{t} \lb\left(\sqrt{2}-16\right) -\sqrt{402-96 \sqrt{2}}\rb,$$ where $$b= -\frac{\sqrt{201-48 \sqrt{2}}}{8} \frac{J}{t} +\sqrt{2}-\frac{3}{8}.$$

We can understand the symmetries of the ground states of this Hamiltonian in the different sectors as follows. For large positive $W$, when the ground states are $\ket{0,2}$ and $\ket{0,4}$, the $\ket{4}$ Hilbert space is one-dimensional with state $$\ket{0,4(B_1)}=\ket{\dn\dn\dn\dn},$$ so the $\ket{4}$ state automatically has $d_{x^2-y^2}$ symmetry. In the limit of infinite $W$, the $\ket{2}$ ground state is a superposition of states with two down spins next to each other. In this limit, the states in the $A_2$ and $B_2$ representations and the two states \beqn
\ket{0,2(E_x)} &\propto& \ket{0\dn\dn0} + \ket{\dn00\dn} + \mathcal{O}\lp \frac{t}{W} \rp,\\
\ket{0,2(E_y)} &\propto& \ket{00\dn\dn} + \ket{\dn\dn00} + \mathcal{O}\lp \frac{t}{W} \rp
\eeqn
in the $E$ representation are the ground states. For finite but large $W$, the $\mathcal{O}(t/W)$ terms, proportional to $\ket{0\dn0\dn}$ and $\ket{\dn0\dn0}$, make the $p$-wave symmetric states the ground states.

As $W$ is decreased, the $\ket{2}$ ground state is $s$-wave $\ket{1,1}$ since positive $J_\perp$ and $J_z$ make it favorable to have molecules of opposite spins. For intermediate values of these interaction strengths, the $\ket{4}$ ground state becomes $\ket{1,3}$. With four molecules on a plaquette, there is no tunneling to consider in the energies of the states. Furthermore, since $n_\up$ and $n_\dn$ are fixed, as determined by $J_z$, $J_\perp$, and $W$ in this sector, we only have to consider the effects of $J_\perp$ to explain the behavior of $\ket{4}$ in this sector. Consider two molecules of opposite spins on sites $r$ and $r'$. The two eigenstates of the $J_\perp$ interaction are $\ket{\pm}_{rr'}$ defined in \eref{pm-def}. Since $J_\perp>0$, the singlet $\ket{-}_{rr'}$ is the ground state of the $J_\perp$ interaction. Therefore, we may expect the $\ket{1,3}$ ground state to be a superposition of singlets on nearest-neighbor bonds with relative phases chosen for constructive interference: $$\lp \dagg{s}_{12}\dagg{t}_{34}+ \dagg{s}_{23}\dagg{t}_{41} + \dagg{s}_{34}\dagg{t}_{12} + \dagg{s}_{41}\dagg{t}_{23}\rp \ket{0}.$$ Here $$t_{rr'}=\dagg{c}_{r\dn} \dagg{c}_{r'\dn}$$ creates the $m=-1$ triplet state on sites $r$ and $r'$, while $s_{rr'}$, defined in \eref{singlet-def}, creates the singlet state on sites $r$ and $r'$. This state is indeed the $\ket{1,3}$ ground state and has the $s$-wave $A_2$ symmetry.

\bibliography{/home/kuns/Documents/Books/papers,/home/kuns/Documents/Books/books-hard,/home/kuns/Documents/Books/books-electronic}

\end{document}